\documentclass[aos,preprint]{imsart}
\RequirePackage[OT1]{fontenc}
\RequirePackage{amsthm,amsmath}

\usepackage{natbib,color,amssymb,amsmath,graphicx,booktabs,amsthm,bm,multirow,threeparttable,textcomp,caption,subcaption}
\usepackage{soul,mathtools}
\usepackage[capposition=top]{floatrow}
\usepackage{xr}
\usepackage{tablefootnote}

\usepackage{threeparttable}

\newcommand\independent{\protect\mathpalette{\protect\independenT}{\perp}}
\def\independenT#1#2{\mathrel{\rlap{$#1#2$}\mkern2mu{#1#2}}}
\makeatletter
\newcommand*{\indep}{%
\mathbin{%
\mathpalette{\@indep}{}%
}%
}
\newcommand*{\nindep}{%
\mathbin{% % The final symbol is a binary math operator
\mathpalette{\@indep}{\not}% \mathpalette helps for the adaptation
}%
}
\newcommand*{\@indep}[2]{%
\sbox0{$#1\perp\m@th$}% box 0 contains \perp symbol
\sbox2{$#1=$}% box 2 for the height of =
\sbox4{$#1\vcenter{}$}% box 4 for the height of the math axis
\rlap{\copy0}% first \perp
\dimen@=\dimexpr\ht2-\ht4-.2pt\relax
\kern\dimen@
{#2}%
\kern\dimen@
\copy0 % second \perp
} 
\makeatother

\arxiv{arXiv:0000.0000}

\startlocaldefs
\numberwithin{equation}{section}
\theoremstyle{plain}

\endlocaldefs

\theoremstyle{definition}

\newtheorem{proposition}{Proposition}
\newtheorem{condition}{Condition}
\newtheorem{corollary}{Corollary}
\newtheorem{lemma}{Lemma}
\newtheorem{example}{Example}

\def\logit{\text{logit }}
\def\expit{\text{expit }}

\begin{document}

\begin{frontmatter}
\title{Identification and Inference for Marginal Average Treatment Effect on the Treated With an Instrumental Variable}
\runtitle{Identification and Inference for Marginal ETT with an IV}

\begin{aug}
\author{\fnms{Lan} \snm{Liu}\thanksref{t1,m1}\ead[label=e1]{liux3771@umn.edu}},
\author{\fnms{Wang} \snm{Miao}\thanksref{t2,m2,m3}\ead[label=e2]{mwfy@pku.edu.cn}}
\author{\fnms{Baoluo} \snm{Sun}\thanksref{t1,m4}\ead[label=e3]{sunb1@gis.a-star.edu.sg}}
\author{\fnms{James} \snm{Robins}\thanksref{t1,m2}\ead[label=e4]{robins@hsph.harvard.edu}}
\and
\author{\fnms{Eric} \snm{Tchetgen Tchetgen}\thanksref{t1,m2}\ead[label=e5]{etchetge@hsph.harvard.edu}}

\thankstext{t1}{Supported by R01 AI032475, R21 AI113251, R01 ES020337, R01 AI104459}
\thankstext{t2}{Supported by China Scholarship Council}
%\thankstext{t3}{Second supporter of the project}
\runauthor{Liu et al.}

\affiliation{University of Minnesota at Twin Cities\thanksmark{m1},\\ Harvard University\thanksmark{m2}, Peking University\thanksmark{m3}\\ and Genome Institute of Singapore\thanksmark{m4}}

\address{Lan Liu\\
University of Minnesota at Twin Cities,\\
224 Church St SE, \# 313 Ford Hall,\\
 Minneapolis, MN 55455\\
\printead{e1}\\
\phantom{E-mail:\ }}

\address{Wang Miao\\
%Beijing International Center\\ for Mathematical Research\\
Peking University\\
Haidian District, Beijing 100871\\
\printead{e2}\\
}

\address{Baoluo Sun\\
Genome Institute of Singapore\\
60 Biopolis St, \#02-01,\\ Singapore 138672\\
\printead{e3}\\
}

\address{James Robins\\
Harvard University,\\
677 Huntington Avenue\\
Kresge,  \#  822\\
Boston, Massachusetts 02115\\
\printead{e4}\\
}

\address{Eric Tchetgen Tchetgen\\
Harvard University,\\
677 Huntington Avenue\\
Kresge,  \#  822\\
Boston, Massachusetts 02115\\
\printead{e5}\\
}

\end{aug}

\begin{abstract}
In observational studies, treatments are typically not randomized and therefore estimated treatment effects may be subject to confounding bias. The instrumental variable (IV) design plays the role of a quasi-experimental handle since the IV is associated with the treatment and only affects the outcome through the treatment. In this paper, we present a novel framework for identification and inference using an IV for the marginal average treatment effect amongst the treated (ETT) in the presence of unmeasured confounding. For inference, we propose three different semiparametric approaches: (i) inverse probability weighting (IPW), (ii) outcome regression (OR), and (iii) doubly robust (DR) estimation, which is consistent if either (i) or (ii) is consistent, but not necessarily both. A closed-form locally semiparametric efficient estimator is obtained in the simple case of binary IV and outcome and the efficiency bound is derived for the more general case.
\end{abstract}

\begin{keyword}[class=MSC]
\kwd[Primary ]{62G05}
\kwd{62G20}
\kwd[; secondary ]{62G35}
\end{keyword}
  
\begin{keyword}
\kwd{Counterfactuals}
\kwd{Double robustness}
\kwd{Instrumental variable}
\kwd{Unmeasured confounding}
\kwd{Effect of treatment on the treated}
\end{keyword}

\end{frontmatter}

\section{\bf Introduction}
Sociology and epidemiology studies often aim to evaluate the effect of a treatment. For practical reasons, the average treatment effect among treated individuals (ETT) is sometimes of greater interest than the treatment effect in the population. For example, in epidemiology studies concerning the toxic effects of a new drug or in sociology studies evaluating the effects of a policy among those whom the policy is applied to, the ETT is the parameter of interest. 

In observational or randomized studies with non-compliance, a primary challenge is the presence of unmeasured confounding, i.e. outcomes between treatment groups may differ not only due to the treatment effect, but also because of unmeasured factors that may affect the treatment selection. %Such phenomenon is addressed as unmeasured confounding.

Instrumental variables (IV) are useful in addressing unmeasured confounding. An IV is a variable that is associated with the treatment and it affects the outcome only through the treatment. The key idea of the IV method is to extract exogenous variation in the treatment that is unconfounded with the outcome and to take advantage of this bias-free component to make causal inference about the treatment effect \citep{robins1989analysis,angrist1996identification,heckman1997instrumental}. 

%%%%%\vspace{-3mm} 
The development of the IV approach can be traced back to \cite{wright1928appendix} and \cite{goldberger1972structural} under linear structural equations in econometrics. \cite{imbens1994identification}, \cite{angrist1996identification} and \cite{heckman1997instrumental} formalized the IV approach within the framework of potential outcomes or counterfactuals. \cite{robins1989analysis} and \cite{robins1994correcting} evaluated the average treatment effect among treated individuals (ETT) conditional on the IV and observed covariates under additive and multiplicative structural nested models (SNMs). Identification is achieved by assuming a certain degree of homogeneity with regard to the IV in an SNM of the conditional ETT \citep{hernan2006instruments}. Mainly, the assumption states that the magnitude of the conditional ETT does not vary with the IV. This is also referred to as the no-current treatment value interaction assumption. Under a similar identifying assumption, \cite{vansteelandt2003causal}, \cite{robins2004estimation}, \cite{tan2010marginal}, \cite{clarke2013estimating} and \cite{Matsouaka2014likelihood} investigated estimation of this conditional causal effect using additive, multiplicative and logistic SNMs.\footnote{In another line of research, \cite{imbens1994identification} and \cite{angrist1996identification} defined the treatment effect on individuals who would comply to their assigned treatment. Under a monotonicity assumption about the effect of the IV on exposure, the complier average treatment effect can be identified. Further research along these lines include fully parametric estimation strategies \citep{tan2006regression,barnard2003principal,frangakis2004methodology} as well as semiparametric methods \citep{abadie2003semiparametric,abadie2002instrumental,tan2006regression,ogburn2014DR}.}

%However, the so-called compliers are themselves not individually identified since only one of the two potential treatments can be observed and the monotonicity assumption will often not be credible. 
%Similar identification results are developed by \cite{joffe2003weighting} in the context of a structural distribution model. 
%%%%%\vspace{-4mm}
The literature mentioned above has some limitations. First of all, the literature focuses on the ETT conditional on the IV and observed covariates. The identification of such conditional ETT was achieved by specifying a functional form of the treatment causal effect. This is unattractive since it places constraints directly on the main parameter of interest and the misspecification of this functional form would lead to biased result. Second, the available inference methods require the treatment propensity score to be correctly specified even for an outcome regression-based estimator \citep{tan2010marginal}. %proposed inverse probability weighting (IPW), outcome regression (OR) and doubly robust (DR) estimator for the conditional ETT. However, the regression estimator he proposed unintuitively requires models for both the treatment propensity score and the outcome regression function to be correctly specified. 

In this paper, we remedy these limitations in a novel framework for identification and estimation using an IV of the marginal ETT in the presence of unmeasured confounding. By targeting directly the marginal ETT, we allow the conditional causal effect to remain unrestricted. Our methods are particularly valuable when the primary goal is to obtain an accurate estimate of the treatment effect. Additionally, we propose a new identification strategy which is applicable to any type of outcome, and provides necessary and sufficient global identification conditions. Moreover, for inference, we propose three different semiparametric estimators allowing for flexible covariate adjustment, (i) inverse probability weighting (IPW), (ii) outcome regression (OR) and (iii) doubly robust (DR) estimation which is consistent if either (i) or (ii) is consistent but not necessarily both.

The outline for the paper is as follows. In Section \ref{sec: preliminaries}, we introduce the notation and state the main assumptions. We study the nonparametric identification of ETT in Section \ref{sec: identification}. We introduce IPW, OR as well as DR estimators in Section \ref{sec: inferences}. In Section \ref{sec: simulations}, we assess the performance of various estimators in a simulation study. In Section \ref{sec: application}, we further illustrate the methods with a study concerning the impact of participation in a 401(k) retirement programs on savings. We conclude with a brief discussion in Section \ref{sec: dis}.

\section{\bf Preliminary Results}\label{sec: preliminaries}
Suppose that one observes independently and identically distributed data $O=(A,Y,Z,C)$, where $A$ is a binary treatment, $Y$ is the outcome of interest and $(Z,C)$ are pre-exposure variables. Let $a,y,z,c$ denote the possible values that $A,Y,Z,C$ could take. Let $Y_{az}$ denote the potential outcome if $A$ and $Z$ are set to $a$ and $z$ and let $Y_{a}$ denote the potential outcome only $A$ is set to $a$. We formalize the IV assumptions using potential outcomes:\\
(IV.1) Stochastic exclusion restriction:
\[Y_{az}=Y_a\text{ almost surely for all }a \text{ and }z;\]
(IV.2) Unconfounded IV-outcome relation:
$$f_{Y_{0}|Z,C}(y|z,c)=f_{Y_{0}|C}(y|c)\text{ for all }z \text{ and }c;$$ 
(IV.3) IV relevance:
$$\Pr(A=1|Z=z,C=c)\neq\Pr(A=1|Z=0,C=c)\text{ for all $z\neq0$ and $c$}.$$

Assumption (IV.1) states that $Z$ does not have a direct effect on the outcome $Y$ thus we use $Y_a$ to denote the potential outcome under treatment $a$ for $a=0,1$. Assumption (IV.2) is ensured under physical randomization but will hold more generally if $C$ includes all common causes of $Z$ and $Y$. Assumptions (IV.1)--(IV.2) together imply that conditional on $C$, the IV is independent of the potential outcome for the unexposed, i.e., $Y_0\independent Z|C$. Assumption (IV.3) states that $A$ and $Z$ have a non-null association conditional on $C$, even if the association is not causal. If assumptions (IV.1)--(IV.3) are satisfied, $Z$ is said to be a valid IV. 

We make the consistency assumption $Y=AY_1+(1-A)Y_0$ almost surely. The marginal treatment effect on the treated is $\text{ETT}=E(Y_1-Y_0|A=1)$. Because $E(Y_1|A=1)=E(Y|A=1)$ can be consistently estimated from the average observed outcome of treated individuals, throughout, we focus on making inferences about $\psi$ where
\begin{equation*}\label{eq: def_psi}
\psi=E(Y_0|A=1).
\end{equation*}

Suppose there exist unmeasured variables denoted by $U$ such that controlling for $(U,Z,C)$ suffices to account for confounding, i.e. $Y_0\indep A|(U,Z,C)$, however, 
\begin{equation}\label{eq: indep_A_Y0}
Y_0\nindep A|(Z,C),
\end{equation}
where $\indep$ denotes statistical independence. As pointed out by \cite{robins2000sensitivity}, potential outcomes can be viewed as the ultimate unmeasured confounders. This is because by the consistency assumption, the observed outcome $Y$ is a deterministic function of the treatment and the potential outcomes. Thus, given $(Y_0,Y_1)$, $U$ does not contain any further information about $Y$. To make explicit use of \eqref{eq: indep_A_Y0}, we define the extended propensity score $\pi(Y_0,Z,C)=\Pr(A=1|Y_0,Z,C)$ as a function of $Y_0$.

%\vspace{-3mm}
\section{\bf Nonparametric Identification}\label{sec: identification}
%In this section, we first consider the case of binary variables. 
%Here we give necessary and sufficient conditions for identification of the joint distribution of $(A,Y_0,Z,C)$. We also give a sufficient condition for identification of marginal ETT which is easier to check in practice.
%These conditions are further illustrated with examples.

While assumptions (IV.1)--(IV.3) suffice to obtain a valid test of the sharp null hypothesis of no treatment effect \citep{robins1994correcting} and can also be used to test for the presence of confounding bias \citep{pearl1995testability}, ETT is not uniquely determined by the observed data without any additional restriction. For simplicity, we first consider the situation where covariates are omitted and outcome and IV are both binary. From the observed data, one can identify the quantities $\Pr(Y_0,Z|A=0)$, $\Pr(Z|A=1)$ and $\Pr(A=0)$. These quantities are functions of the unknown parameters: $\Pr(Z=1)$, $\Pr(Y_0=1)$, and $\Pr(A=0|Y_0,Z)$. Without imposing any additional assumption, there are six unknown parameters (one for $\Pr(Z=1)$, one for $\Pr(Y_0=1)$ and four for $\Pr(A=0|Y_0,Z)$), however, only five degrees of freedom are available from the observed data (one for $\Pr(A=0)$, one for $\Pr(Z|A=0)$ and three for $\Pr(Y,Z|A=0)$). As a result, the joint distribution $f(A,Y_0,Z)$ is not uniquely identified. Particularly, $\psi$ is not identified. 

For identification purposes, additional assumptions, such as Robins' no-current treatment value interaction assumption \citep{hernan2006instruments}, must be imposed to reduce the set of candidate models for the joint distribution $f(A,Y_0,Z,C)$. Below, we give a general necessary and sufficient condition for identification. 
%We restrict the candidates for the joint distribution to a smaller set, which is a subset of all distributions satisfying assumptions (IV.1)--(IV.3). 
Let $\mathcal{P}_{A|Y_0,Z,C}$ and $\mathcal{P}_{Y_0|C}$ denote the collections of candidates for $\Pr(A=0|Y_0,Z,C)$ and $f(Y_0|C)$, which are known to satisfy (IV.1) and (IV.2).

\begin{condition}\label{cond: not1}
Any two distinct elements $\Pr_1(A=0|Y_0,Z,C)$, $\Pr_2(A=0|Y_0,Z,C)\in \mathcal{P}_{A|Y_0,Z,C}$ and $f_1(Y_0|C)$, $f_2(Y_0|C)\in \mathcal{P}_{Y_0|C}$, satisfy the inequality:
\begin{equation*}\label{eq: not1}
\frac{\Pr_1(A=0|Y_0,Z,C)}{\Pr_2(A=0|Y_0,Z,C)}\neq \frac{f_2(Y_0|C)}{f_1(Y_0|C)}.
\end{equation*}
\end{condition}

The following proposition states that condition \ref{cond: not1} is a necessary and sufficient condition for identifiability of the joint distribution of $(A,Y_0,Z,C)$, where $Y_0$ and $Z$ may be dichotomous, polytomous, discrete or continuous.

\begin{proposition}\label{prop: indenti}
The joint distribution of $(A,Y_0,Z,C)$ is identified in the model defined by $\mathcal{P}_{A|Y_0,Z,C}$ and $\mathcal{P}_{Y_0|C}$ if and only if condition \ref{cond: not1} holds. 
\end{proposition}

%Note that although (IV.3) is needed in the definition of IV, it is not strictly required for identification. 
It is convenient to check condition \ref{cond: not1} for parametric models, but it may be harder for semiparametric and nonparametric models, since $\mathcal{P}_{A|Y_0,Z,C}$ and $\mathcal{P}_{Y_0|C}$ can be complicated. The following corollary gives a more convenient condition.

\begin{corollary}\label{coro: indenti}
Suppose that for any two candidates $\Pr_1(A=0|Y_0,Z,C)$, $\Pr_2(A=0|Y_0,Z,C)\in \mathcal{P}_{A|Y_0,Z,C}$, the ratio $\Pr_1(A=0|Y_0,Z,C)/\Pr_2(A=0|Y_0,Z,C)$ is either a constant or varies with $Z$. Then the joint distribution of $(A,Y_0,Z,C)$ is identified.
\end{corollary}

Although the condition provided in Corollary \ref{coro: indenti} is a sufficient condition for identification, it allows identification of a large class of models. We further illustrate Proposition \ref{prop: indenti} and Corollary \ref{coro: indenti} with several examples. For simplicity, we again omit covariates, however, we show at the end of this section that similar results with covariates can be derived. We first consider the case of binary outcome with binary IV.

\begin{example}\label{eg: saturate_logit} 
Consider a model 
$\mathcal{P}_{A|Y_0,Z}=\{\Pr(A=0|Y_0,Z):\logit\Pr(A=0|Y_0,Z;\theta_1,\theta_2,\eta_1,\eta_2)=\theta_1+\theta_2 Z+\eta_1 Y_0+\eta_2 Y_0Z,\theta_1,\theta_2,\eta_1,\eta_2\in(-\infty,\infty)\}.$
The model is saturated since $\mathcal{P}_{A|Y_0,Z}$ contains all possible treatment mechanisms.
It can be shown that neither the joint distribution nor $\psi$ is identified even under the assumptions (IV.1)--(IV.3). 
\end{example}

Example \ref{eg: saturate_logit} shows that the joint density $f(A,Y_0,Z)$ is not identified when the treatment selection mechanism is left unrestricted under (IV.1)--(IV.3). However, we show that the joint density $f(A,Y_0,Z)$ is identified assuming separable treatment mechanism on the additive scale.

\begin{example}\label{eg: sep_logit_bin}
Consider a model
$\mathcal{P}_{A|Y_0,Z}=\{\Pr(A=0|Y_0,Z):\logit\Pr(A=0|Y_0,Z;\theta_1,\theta_2,\eta_1) = \theta_1 +\theta_2Z+\eta_1 Y_0;\theta_1,\theta_2,\eta_1\in(-\infty,\infty)\}.$
The model is separable since $\mathcal{P}_{A|Y_0,Z}$ excludes an interaction between $Y_0$ and $Z$.
It can be shown that both the joint distribution and $\psi$ is identified under assumptions (IV.1)--(IV.3). 
\end{example}

Example \ref{eg: sep_logit_bin} agrees with the intuition that identification follows from having fewer parameters than the saturated model. Under the assumed model, we have five unknown parameters and five available degrees of freedom from the empirical distribution. We show in the next example that the joint distribution and $\psi$ can be identified in a general separable model when the outcome and instrument are both continuous. 

%The IV model with separable treatment mechanism is also identified for continuous outcome with continuous instrument.
\begin{example}\label{eg: sep}
Consider the logistic separable treatment mechanism: $\mathcal{P}_{A|Y_0,Z}\\=\{\Pr(A=0|Y_0,Z):\logit\Pr(A=0|Y_0,Z) = q(Z) + h(Y_0)\}$, where $q$ and $h$ are unknown differentiable functions with $h(0)=0$. It can be shown that $\mathcal{P}_{A|Y_0,Z}$ satisfies condition \ref{cond: not1} and thus the joint distribution is identified under (IV.1)--(IV.3). 
\end{example}

%however, if $Y_0$ and $Z$ do not interact in the extended propensity score function, the joint distribution is identified as shown in examples \ref{eg: saturate_logit} and \ref{eg: sep}.

%While the model in example \ref{eg: sep} rules out the presence of an interaction between $Y_0$ and $Z$ on the logit scale, it does not a priori rule the presence of such interaction on any other scale, say additive or multiplicative. 
These results can be generalized to include covariates $C$. For instance, by allowing both $q$ and $h$ to depend on $C$ in example \ref{eg: sep}:
\[\mathcal{P}_{A|Y_0,Z,C}=\{\Pr(A=0|Y_0,Z,C):\logit \Pr(A=0|Y_0,Z,C) = q(Z,C) + h(Y_0,C)\},\]

\noindent 
where $h(0,C)=0$, the joint distribution is identified whenever the interaction term of $Y_0$ and $Z$ is absent. 

In the Supplementary Materials, we present proofs for the above examples, and additional examples, such as the case of continuous outcome with binary IV, and a separable treatment mechanism. 
%\vspace{-5mm}
\section{\bf Estimation}\label{sec: inferences}
While nonparametric identification conditions are provided in Section \ref{sec: identification}, such conditions will seldom suffice for reliable statistical inference. Typically in observational studies, the set of covariates $C$ is too large for nonparametric inference, due to the curse of dimensionality \citep{robins1997toward}. To make progress, we posit parametric models for various nuisance parameters, and provide three possible approaches for semiparametric inference that depend on different subsets of models. We describe an IPW, an OR and a DR estimator of the marginal ETT under assumptions (IV.1)--(IV.2) and condition \ref{cond: not1}. Throughout, we posit a parametric model $f_{Z|C}(z|c)=\Pr(Z=z|C=c;\rho)$ for the conditional density of $Z$ given $C$. Let $\hat\rho$ denote the maximum likelihood estimator (MLE) of $\rho$. Let $\mathbb{P}_n$ denote the empirical measure, that is $\mathbb{P}_nf(O)=n^{-1}\sum_{i=1}^n f(O_i)$. Let $\hat E$ denote the expectation taken under the empirical distribution of $C$ and let $\widehat{\Pr}(A=1)=\sum_{i=1}^nA_i/n$ denote the empirical probability of receiving treatment.

%we assume the model specified for the selection bias function $\alpha(Y_0,Z,C;\eta)$ is correctly specified, with $\alpha(Y_0=0,Z,C;\eta)=\alpha(Y_0,Z,C;0)=0$. Also throughout, 

%\vspace{-3mm}
\subsection{IPW estimator}\label{subsec: ipw}
For estimation, we first propose an IPW IV approach which extends standard IPW estimation of ETT to an IV setting. We make the positivity assumption that for all values of $Y_0$, $Z$ and $C$ the probability of being unexposed to treatment is bounded away from 0. The IPW approach relies on the crucial assumption that the extended propensity score model $\pi(Y_0,Z,C;\gamma)$ is correctly specified with unknown finite dimensional parameter $\gamma$ and the following representation of ETT,

%\vspace{-6mm}
\begin{equation}\label{eq: ipw_identity}
E(Y_0|A=1)=E\left\{\frac{{\pi}(Y_0,Z,C)Y(1-A)}{\Pr(A=1)\{1-{\pi}(Y_0,Z,C)\}}\right\}.
\end{equation} 

%\vspace{-0.5mm}
A derivation of the above equation is given in the Supplementary Materials. We solve the following equations to obtain an estimator $\hat\gamma$ of $\gamma$:

%\vspace{-9.5mm}
\begin{eqnarray}\label{eq: ipw_intercept}
&&\mathbb{P}_n\{ \frac{1-A}{1-{\pi}(Y_0,Z,C;\hat\gamma)}-1\}=0,\\
%\end{equation}
%\begin{equation}
&&\mathbb{P}_n\bigl[\frac{1-A}{1-{\pi}(Y_0,Z,C;\hat\gamma)}\{h_1(Z,C)- E(h_1(Z,C)|C;\hat\rho)\}\bigr]=0\label{eq: ipw_Z},\\
%\end{equation}
%\begin{equation}
&&\mathbb{P}_n\bigl[\frac{1-A}{1-{\pi}(Y_0,Z,C;\hat\gamma)}\{h_2(C)-\hat E(h_2(C))\}\bigr]=0\label{eq: ipw_C},\\
%\end{equation}
%\begin{equation}
&&\mathbb{P}_n\bigl[\frac{1-A}{1-{\pi}(Y_0,Z,C;\hat\gamma)}t(Y,C)\{l(Z,C)-E(l(Z,C)|C;\hat\rho)\}\bigr]=0,\label{eq: ipw_alpha}
\end{eqnarray}
where $(h_1^T,h_2^T,l^T)^T$ satisfies the regularity condition \eqref{eq: cond_uniqueness_ipw} described in the Supplementary Materials. Equations \eqref{eq: ipw_Z} and \eqref{eq: ipw_C} identify the association between $(Z,C)$ and $A$ in $\pi(0,Z,C)$. By leveraging the IV property (IV.1)--(IV.2), equation \eqref{eq: ipw_alpha} identifies the degree of selection bias encoded in the dependence of $\pi$ on $Y_0$. By equation \eqref{eq: ipw_identity}, an extended propensity score estimator  leads to an estimator of $\psi$. We have the following result:

\begin{proposition}\label{prop: ipw_unbias}
Under (IV.1)--(IV.2) and condition \ref{cond: not1}, suppose the extended propensity score model $\pi(Y_0,Z,C;\gamma)$ and $f_{Z|C}(z|c;\rho)$ are correctly specified, then the IPW estimator
\begin{equation*}
\hat{\psi}^{ipw}=\mathbb{P}_n\frac{{\pi}(Y_0,Z,C;\hat\gamma)Y(1-A)}{\widehat\Pr(A=1)\{1-{\pi}(Y_0,Z,C;\hat\gamma)\}},
\end{equation*}
is consistent for $\psi$.
\end{proposition}

%We emphasize that the extended propensity score can use any well-defined link function (e.g., logit, probit), and Proposition \ref{prop: ipw_unbias} still holds, provided condition \eqref{eq: cond_uniqueness_ipw} in the Appendix holds so that the population expectation value of the derivative of the vector of equations \eqref{eq: ipw_intercept}--\eqref{eq: ipw_alpha} is invertible when evaluated at the true parameter value. 

We emphasize that the extended propensity score model can use any well-defined link function (e.g., logit, probit), and if condition \ref{cond: not1} holds, Proposition \ref{prop: ipw_unbias} still holds. The functions $h_1$, $h_2$, $t$ and $l$ can be chosen based on the model for the extended propensity score. For example, assuming $\logit\pi(Y_0,Z,C;\gamma)=\theta_0+\theta_1Z+\theta_2C+\eta Y_0$ where $\tilde{\eta}=(\theta_1,\theta_2,\eta)^T$ is a $k$-dimensional parameter vector. The $k$-dimensional function $(h_1,h_2,t)^{T}$ can be chosen as $(h_1,h_2,t)^T=\partial \logit \pi(Y_0,Z,C;\gamma)/\partial \tilde{\eta}=(Z,C,Y_0)^T$ and $l$ can be chosen as any scalar function of $(Z,C)$, e.g., $l(Z,C)=Z$. Thus we have exactly $k+1$ estimating equations. The choice of $h_1$, $h_2$, $t$ and $l$ will generally impact efficiency but should not affect consistency as long as the identification conditions hold and the required models are correctly specified.

% For simplicity, we assume that $\kappa$ and $\lambda$ are both the logit link unless otherwise stated.
%the conditional mean of $Y_0$ among treated individuals given pre-treatment covariates
\subsection{OR and DR estimators}\label{subsec: reg_DR}

Since $Y_0$ is never observed for the treated group, we parameterize $E[Y_0|A=1,Z,C]$ into two parts: one can be estimated directly using restricted MLE and the other can be computed by solving an estimating equation. Specifically, we have 

\begin{equation}\label{eq: moment_eq}
%E(Y_0|A=1,Z,C)=\frac{E[\exp\{\alpha(Y,Z,C)\}Y|A=0,Z,C]}{E[\exp\{\alpha(Y,Z,C)\}|A=0,Z,C]}.
E\{g(Y_0,C)|A=1,Z,C\}=\frac{E[\exp\{\alpha(Y,Z,C)\}g(Y,C)|A=0,Z,C]}{E[\exp\{\alpha(Y,Z,C)\}|A=0,Z,C]},
\end{equation}

\noindent where $g$ is any function of $Y_0$ and $C$ and  $\alpha(Y_0,Z,C)$ is the generalized odds ratio function relating $A$ and $Y_0$ conditional on $Z$ and $C$ as 
 
\begin{equation*}
\alpha(Y_0,Z,C)=\log\frac{f(Y_0|A=1,Z,C)f(Y_0=0|A=0,Z,C)}{f(Y_0|A=0,Z,C)f(Y_0=0|A=1,Z,C)}.
\end{equation*}

\noindent Since the association between $Y_0$ and $A$ is attributed to unmeasured confounding, $\alpha(Y_0,Z,C)$ can be interpreted as the selection bias function. Thus, we express the conditional mean function $E\{g(Y_0,C)|A=1,Z,C\}$ in terms of $f(Y|A=0,Z,C)$ and $\alpha(Y_0,Z,C)$. We prove the equation \eqref{eq: moment_eq} in the Supplementary Materials.

Let $f(Y|A=0,Z,C;\xi)$ denote a model for the density of the outcome among the unexposed conditional on $Z$ and $C$, and let $\hat\xi$ denote the restricted MLE of $\xi$ obtained using only data among the unexposed. Let $\eta$ denote the parameter indexing a parametric model for the selection bias function $\alpha$ as $\alpha(Y_0,Z,C;\eta)$. We obtain an estimator for $\eta$ by solving:
%\begin{equation}\label{eq: reg_Y0}
%\mathbb{P}_n \bigl\{Y-E(Y|A=0,Z,C)\bigr\}(1-A)u(Z,C)=0
%\end{equation}
\begin{equation}\label{eq: reg_alpha}
%\mathbb{P}_n\biggl[\bigl\{w(Z,C)- E(w(Z,C)|C;\hat\rho)\bigr\}\biggl\{A\frac{ E[\exp\{\hat\alpha(Y,Z,C;\eta)\}g(Y,C)|A=0,Z,C;\hat\xi]}{ E[\exp\{\hat\alpha(Y,Z,C;\eta)\}|A=0,Z,C;\hat\xi]}+(1-A)g(Y,C)\biggr\}\biggr]=0,
\mathbb{P}_n\biggl[\bigl\{w(Z,C)- E(w(Z,C)|C;\hat\rho)\bigr\}\biggl\{AE[g(Y_0,C)|A=1,Z,C;\eta,\hat\xi]+(1-A)g(Y,C)\biggr\}\biggr]=0,
\end{equation}
for any choice of functions $w$ and $g$ such that the regularity condition \eqref{eq: cond_uniqueness_reg} stated in the Supplementary Materials holds. Intuitively, the left hand side of equation \eqref{eq: reg_alpha} is an empirical estimator of the expected conditional covariance between $g(Y_0,C)$ and $w(Z,C)$ given C, which should be zero by (IV.1)--(IV.2). Based on equation \eqref{eq: moment_eq}, we can construct an estimator for $\psi$ based on $\hat\eta$, $\hat\xi$ and $\hat{\rho}$.
\begin{proposition}\label{prop: reg_unbias}
Under (IV.1)--(IV.2) and condition \ref{cond: not1}, suppose $\alpha(Y_0,Z,C;\eta)$, $f_{Z|C}(z|c;\rho)$ and $f(Y|A=0,Z,C;\xi)$ are correctly specified, then the OR estimator 
\begin{equation*}
\hat{\psi}^{reg}=\mathbb{P}_n\frac{A}{\hat\Pr(A=1)}\frac{{E}[\exp\{\alpha(Y,Z,C;\hat\eta)\}Y|A=0,Z,C;\hat\xi]}{{E}[\exp\{\alpha(Y,Z,C;\hat\eta)\}|A=0,Z,C;\hat\xi]},
\end{equation*}
is consistent for $\psi$.
\end{proposition}

Functions $g$ and $\omega$ in equation \eqref{eq: reg_alpha} can be chosen based on the model we posit for $\alpha(Y_0,Z,C)$. For example, assuming 
\begin{equation}\label{eq: alpha=Y0}
\alpha(Y_0,Z,C;\eta)=\eta Y_0,
\end{equation}
$g$ can be chosen as $g(Y_0,C)=\partial \alpha(Y_0,Z,C;\eta)/\partial \eta=Y_0$ and $\omega$ can be chosen as any scalar function of $(Z,C)$, e.g., $\omega(Z,C)=Z$. The choice of $g$ and $\omega$ may impact efficiency but does not affect consistency as long as the identification conditions hold and the required models are correctly specified.

\cite{tan2010marginal} proposed an OR estimator for the conditional ETT, which requires correctly specified models for both the treatment propensity score and the outcome regression function. In contrast, we circumvent the dependence of the regression estimator on the propensity score.

%{\blue To emphasis, I recommend adding a theorem for selection bias being DR. it worths a theorem, that's the key for dr for ETT}

Note that the proposed estimator for nuisance parameter $\eta$ is closely related to the regression estimator proposed by \cite{vansteelandt2003causal} when $Y$ is binary. \cite{vansteelandt2003causal} developed a two-stage logistic estimator which combines a logistic SMM at the first stage and a logistic regression association model at the second stage. Specifically, \cite{vansteelandt2003causal} focused on estimating
$\zeta(Z,C)=\logit \Pr(Y_1=1|A=1,Z,C)-\logit \Pr(Y_0=1|A=1,Z,C)$, which encodes the conditional ETT given $Z$ and $C$. Let $\nu$ denote the parameter indexing a model for $\zeta(Z,C)$ as $\zeta(Z,C;\nu)$. They proposed to estimate $\nu$ in the estimating equation

\begin{equation}\label{eq: reg_alpha_stjin}
\mathbb{P}_n\biggl\{\bigl(w(Z,C)- E(w(Z,C)|C;\hat\rho)\bigr)\biggl(A\text{expit}\{\vartheta(Z,C;\hat\varrho)-\zeta(Z,C;\nu)\}+(1-A)Y\biggr)\biggr\}=0.
\end{equation}
%{\blue brackets are not right}

\noindent where $\text{expit}(x)=\exp (x)/\{1+\exp (x)\}$ and $\vartheta(Z,C;\varrho)=\logit \Pr(Y=1|A=1,Z,C;\varrho)$.

Recall that we obtain an estimator of $\eta$ indexing $\alpha(Y_0,Z,C;\eta)$ in the equation \eqref{eq: reg_alpha}, which can be re-expressed as
\begin{equation}\label{eq: reg_alpha_intermsof_Y}
\mathbb{P}_n\biggl\{\bigl(w(Z,C)- E(w(Z,C)|C;\hat\rho)\bigr)\biggl(A\text{expit}\{\delta(Z,C;\hat\xi)+\alpha(1,Z,C;\eta)\}+(1-A)Y\biggr)\biggr\}=0,
\end{equation}
%{\blue brackets are not right}
where $\delta(Z,C;\hat\xi)=\logit \Pr(Y_0=1|A=0,Z,C)$. Equations \eqref{eq: reg_alpha_stjin} and \eqref{eq: reg_alpha_intermsof_Y} mainly differ in the way $\Pr(Y_0=1|A=1,Z,C)$ is estimated. More specifically, \eqref{eq: reg_alpha_stjin} obtains $\Pr(Y_0=1|A=1,Z,C)$ using $\Pr(Y_1=1|A=1,Z,C)$ as a baseline risk for the model while \eqref{eq: reg_alpha_intermsof_Y} uses $\Pr(Y_0=1|A=0,Z,C)$ as baseline risk. This difference is important since \cite{vansteelandt2003causal} failed to obtain a DR estimator of $\zeta(Z,C)$ while as we show next, our choice of parameterization yields a DR estimator of the marginal ETT. 

%\subsection{{\color{blue}doubly} robust estimator}\label{subsec: reg_DR}
%Let $\beta(Z,C)=\logit\Pr(A=1|Y_0=0,Z,C)$ denotes the baseline extended propensity score. Thus, under the logit link for the treatment extended propensity, we have $\logit{\pi}(Y_0,Z,C;\gamma)=\alpha(Y_0,Z,C;\eta)+\beta(Z,C;\theta)$, where $\gamma=(\eta,\theta)$. 

Heretofore, we have constructed estimators in two different approaches. Both approaches assume correct models for $\alpha(Y_0,Z,C;\eta)$ and $f_{Z|C}(z|c;\rho)$. The IPW approach further relies on a consistent estimator of the baseline extended propensity score $\beta(Z,C)=\logit\Pr(A=1|Y_0=0,Z,C)$, which under the logit link and together with $\alpha(Y_0,Z,C;\eta)$, provides a consistent estimator of the extended propensity score $\pi(Y_0,Z,C;\gamma)=\expit\{\alpha(Y_0,Z,C;\eta)+\beta(Z,C;\theta)\}$. The OR approach further relies on a consistent estimator of $f(Y|A=0,Z,C)$, which together with $\alpha(Y_0,Z,C;\eta)$, provides a consistent estimator of $\Pr(Y_0=1|A=1,Z,C)$ by \eqref{eq: moment_eq}. Define $\mathcal{M}_a$ as the collection of laws with parametric models $f_{Z|C}(z|c;\rho)$, $\alpha(Y_0,Z,C;\eta)$ and $\beta(Z,C;\theta)$ while $f(Y|A=0,Z,C)$ is unrestricted. Likewise, define $\mathcal{M}_y$ as the collection of laws with parametric models $f_{Z|C}(z|c;\rho)$, $\alpha(Y_0,Z,C;\eta)$ and $f(Y|A=0,Z,C;\xi)$ while $\beta(Z,C)$ is unrestricted. The main appeal of a doubly robust estimator is that it remains consistent if either $\beta(Z,C;\theta)$ or $f(Y|A=0,Z,C;\xi)$ is correctly specified. To derive a DR estimator for $\psi$ in the union space $\mathcal{M}_a\cup \mathcal{M}_y$, we first propose a DR estimator for the parameter $\eta$ of the selection bias model $\alpha(Y_0,Z,C;\eta)$. For notational convenience, let 
\begin{eqnarray}
&&\label{eq: psi_DR}\\
&&Q_g(Y,A,Z,C;\gamma,\xi)\nonumber\\
&=&\frac{(1-A)\pi(Y,Z,C;\gamma)}{1-\pi(Y,Z,C;\gamma)}\biggl[g(Y,C)-\frac{E[\exp\{\alpha(Y,Z,C;\eta)\}g(Y,C)|A=0,Z,C;\xi]}{E[\exp\{\alpha(Y,Z,C;\eta)\}|A=0,Z,C;\xi]}\biggr]\nonumber\\
&&+A\frac{E[\exp\{\alpha(Y,Z,C;\eta)\}g(Y,C)|A=0,Z,C;\xi]}{E[\exp\{\alpha(Y,Z,C;\eta)\}|A=0,Z,C;\xi]}.\nonumber
\end{eqnarray}
Equation \eqref{eq: psi_DR} is key to obtaining a DR estimation of the selection bias function and thus of ETT. Specifically, consider the estimating equation for the selection bias parameter $\tilde\eta$

\begin{equation}\label{eq: double_robust_selectionbias}
\mathbb{P}_n\biggl[\bigl[\omega(Z,C)- E\{\omega(Z,C)|C;\hat\rho\}\bigr] \tilde Q_g(Y,A,Z,C;\tilde\gamma,\hat\xi)\biggr]=0,
\end{equation}

%\vspace{-5mm}
\noindent where 

\begin{eqnarray*}
&&\tilde Q_g(Y,A,Z,C;\tilde\gamma,\hat\xi)\\
&=&Q_g(Y,A,Z,C;\tilde\gamma,\hat\xi)+(1-A)g(Y,C)\\
&=&\frac{1-A}{1-\pi(Y,Z,C;\gamma)}g(Y,C)\\
&&\hspace{1mm}+\frac{A-\pi(Y,Z,C;\gamma)}{1-\pi(Y,Z,C;\gamma)}\frac{E[\exp\{\alpha(Y,Z,C;\eta)\}g(Y,C)|A=0,Z,C;\xi]}{E[\exp\{\alpha(Y,Z,C;\eta)\}|A=0,Z,C;\xi]}.
\end{eqnarray*}

 We solve equation \eqref{eq: double_robust_selectionbias} jointly with equations \eqref{eq: ipw_intercept}--\eqref{eq: ipw_C} with $\hat \gamma$ replaced by $\tilde\gamma=(\hat\eta^{DR},\tilde\theta)$. The choice of $h_1,h_2,g$ and $w$ can be decided as in Sections 4.1 and 4.2. 
%The proposed DR estimator is based on $\tilde{\gamma}$ and the MLE of $f(Y|A=0,C,Z;\xi)$.

\begin{proposition}\label{prop: DR_unknown_alpha0}
Under (IV.1)--(IV.2) and condition \ref{cond: not1}, $\hat{\eta}^{DR}$ and $\hat{\psi}^{DR}$ are consistent in the union model $\mathcal{M}_{a}\cup \mathcal{M}_{y}$, where $\hat{\psi}^{DR}=\hat{\mathbb{P}}_n Q_{\tilde g}(Y,A,Z,C;\tilde\gamma,\hat\xi)/\\\widehat{\Pr}(A=1)$ and $\tilde g(Y,C)=Y$.
\end{proposition}

Proposition \ref{prop: DR_unknown_alpha0} implies that $\hat{\eta}^{DR}$ and $\hat{\psi}^{DR}$ are both DR estimators since their consistency only requires either the extended propensity score or the outcome regression model to be correctly specified but not necessarily both. 

%\eqref{eq: double_robust_selectionbias}\ref{sec: identification}

%Note that the DR estimator given in proposition \ref{prop: DR_unknown_alpha0} has the properties of doubly robustness, however it is not the most efficient estimator. We derive the local efficiency in the supplementary material. 

%To our knowledge, this is the first DR estimator for ETT that has ever been proposed. 

\subsection{Local efficiency}
The large sample variance of doubly robust estimators $\hat{\eta}^{DR}$ and $\hat{\psi}^{DR}$ at the intersection submodel $\mathcal{M}_{a}\cap \mathcal{M}_{y}$ where all models are correctly specified, is determined by the choice of $g(Y,C)$ and $\omega(Z,C)$ in equation \eqref{eq: double_robust_selectionbias}. In the Supplementary Materials, we derive the semiparametric efficient score of $(\eta,\psi)$ in a model $\mathcal{M}_{np}$ that only assumes that $Z$ is a valid IV and the selection bias function $\alpha(Y_0,Z,C;\eta)$ is correctly specified. As discussed in the Supplementary Materials, the efficient score is generally not available in closed-form, except in special cases, such as when $Z$ and $Y$ are both polytomous. Next, we illustrate the result by constructing a locally efficient estimator of $(\eta,\psi)$ when $Z$ and $Y$ are both binary. In this vein, similar to the definition of $\tilde Q_g(Y,A,Z,C;\gamma,\xi)$, define

%\vspace{-10mm}
\begin{eqnarray*}
&&\tilde Q_v(Y,A,Z,C;\gamma,\xi)=\frac{(1-A)v(Y,Z,C)}{1-\pi(Y,Z,C;\gamma)}\nonumber\\
&&\hspace{2mm}+\frac{A-\pi(Y,Z,C;\gamma)}{1-\pi(Y,Z,C;\gamma)}\frac{E[\exp\{\alpha(Y,Z,C;\eta)\}v(Y,Z,C)|A=0,Z,C;\xi]}{E[\exp\{\alpha(Y,Z,C;\eta)\}|A=0,Z,C;\xi]},
\end{eqnarray*}

\noindent where $v$ is any function of $(Y_0,Z,C)$. 

A one-step locally efficient estimator of $\eta$ in $\mathcal{M}_{np}$ is given by

\[\hat\eta^{eff}=\hat\eta^{DR}-\{E(\bigtriangledown_{\eta}\widehat S^{eff}_{\eta}|\hat\gamma,\hat\xi)\}^{-1}E(\widehat S^{eff}_{\eta}|\hat\gamma,\hat\xi),\]

\noindent where $ \bar v(Y,Z,C)=\{Y-E(Y|C)\}\{Z-E(Z|C)\}$, $\Delta(\eta)=\tilde Q_{\bar v}(Y,A,Z,C;\gamma,\xi)$ and $\widehat S^{eff}_{\eta}={E}\{\Delta(\eta)\Delta(\eta)^T|C;\hat\gamma,\hat\xi\}^{-1}{E}\{\partial \Delta(\eta)/\partial\eta^T|C;\hat\gamma,\hat\xi\}\Delta(\hat\eta^{eff})$ is the efficient score of $\eta$ evaluated at the estimated intersection submodel $\mathcal{M}_{a}\cap \mathcal{M}_{y}$. Further, let $\hat{\psi}^{DR}(\hat\eta^{eff})$ denote a DR estimator for $\psi$ evaluated
at the estimated intersection submodel $\mathcal{M}_{a}\cap \mathcal{M}_{y}$ with $\hat\eta^{eff}$ substituted in for $\hat\eta^{DR}$. Then the efficient estimator of $\psi$ is given by

\[\hat\psi^{eff}=\hat{\psi}^{DR}(\hat\eta^{eff})-E\{\Delta^2(\hat\eta^{eff})|C;\hat\gamma,\hat\xi\}^{-1}E\{\hat{\psi}^{DR}(\hat\eta^{eff})\Delta(\hat\eta^{eff})|C;\hat\gamma,\hat\xi\}\Delta(\hat\eta^{eff}).\]

\section{\bf Simulations}\label{sec: simulations}
Simulations for both binary and continuous outcomes were conducted to evaluate the finite sample performance of the causal effect estimators derived in Sections \ref{subsec: ipw} and \ref{subsec: reg_DR}. Let $\mathcal{M}_a^c$ denote the complement space of $\mathcal{M}_a$ and likewise define $\mathcal{M}_y^c$. Simulations were conducted under three scenarios: (i) $\mathcal{M}_a\cap \mathcal{M}_y$, that is both outcome regression and extended propensity score are correctly specified, (ii) $\mathcal{M}_a\cap \mathcal{M}_y^c$ that is only the extended propensity score is correctly specified and (iii) $\mathcal{M}_a^c\cap \mathcal{M}_y$ that is only the outcome regression model is correctly specified.

Simulations were first carried out for a binary outcome. For scenario (i), the simulation study was conducted in the following steps:
\begin{enumerate}
\item [Step 1:] A hypothetical study population of size $n$ was generated and each individual had baseline covariates $C_1$ and $C_2$ generated independently from Bernoulli distributions with probability 0.4 and 0.6 respectively. Then the IV $Z$ was generated from the model: $\logit \Pr(Z=1|C)=0.2+0.4C_1-0.5C_2$ and potential outcomes $Y_0,Y_1$ from models $\logit \Pr(Y_0\\=1|Z,C)=0.6+0.8C_1-2C_2$ and $\logit \Pr(Y_1=1|Z,C)=0.7-0.3C_1$. The treatment variable $A$ was generated from $\logit \Pr(A=1|Y_0,Z,C)=0.4+2Z+0.8C_1-0.6Y_0-1.6C_1Z$, and the observed outcome was $Y=Y_0(1-A)+Y_1A$.

\item [Step 2:] The following extended propensity score model was estimated and the parameters $\gamma=(\theta_1,\theta_2,\theta_3,\theta_4,\eta)$ in the model 
\begin{equation}\label{eq: propen_correct}
\logit\Pr(A=1|Y_0,Z,C;\gamma)=\theta_1+\theta_2Z+\theta_3C_1+\theta_4C_1Z+\eta Y_0
%\logit \Pr(A=1|Y_0,Z,C)=\gamma_1+\gamma_2Z+\gamma_3C_1+\gamma_4Y_0+\gamma_5C_1Z
\end{equation}

were estimated using estimating equations \eqref{eq: ipw_intercept}--\eqref{eq: ipw_alpha} with $h_1(Z,C)=(Z,C_1Z)^{T}$, $h_2(C)=C_1$, $t(Y,C)=Y$ and $l(Z,C)=Z$ and $\hat{\psi}^{ipw}$ was evaluated.

\item [Step 3:] The selection bias function was correctly specified as in \eqref{eq: alpha=Y0}, $\xi$ in the regression outcome model 

\begin{equation}\label{eq: Y0_correct}
\logit E(Y|A=0,Z,C;\xi)=\xi_1+\xi_2C_1+\xi_3C_2+\xi_4Z+\xi_5C_1Z
\end{equation}

\noindent was estimated by restricted MLE, and $\alpha$ was estimated by solving equation \eqref{eq: reg_alpha} with $\omega(Z,C)=Z$ and $g(Y,C)=Y$ and $\hat{\psi}^{reg}$ was evaluated.
\item [Step 4:] The selection bias function was correctly specified as in \eqref{eq: alpha=Y0}, $\xi$ in equation \eqref{eq: Y0_correct} was estimated by restricted MLE, parameters $\gamma$ in \eqref{eq: propen_correct} was estimated using \eqref{eq: ipw_intercept}--\eqref{eq: ipw_C} and \eqref{eq: double_robust_selectionbias} where $h,t,l,\omega,g$ are chosen as in Step 2 and Step 3 and $\hat{\psi}^{DR}$ was evaluated. 

\item [Step 5:] Steps 1--4 were repeated 1000 times.

\end{enumerate} 

\begin{figure}
\centering
\begin{subfigure}[b]{.47\textwidth}
%  \centering
  \includegraphics[width=.9\linewidth,height=4.5cm]{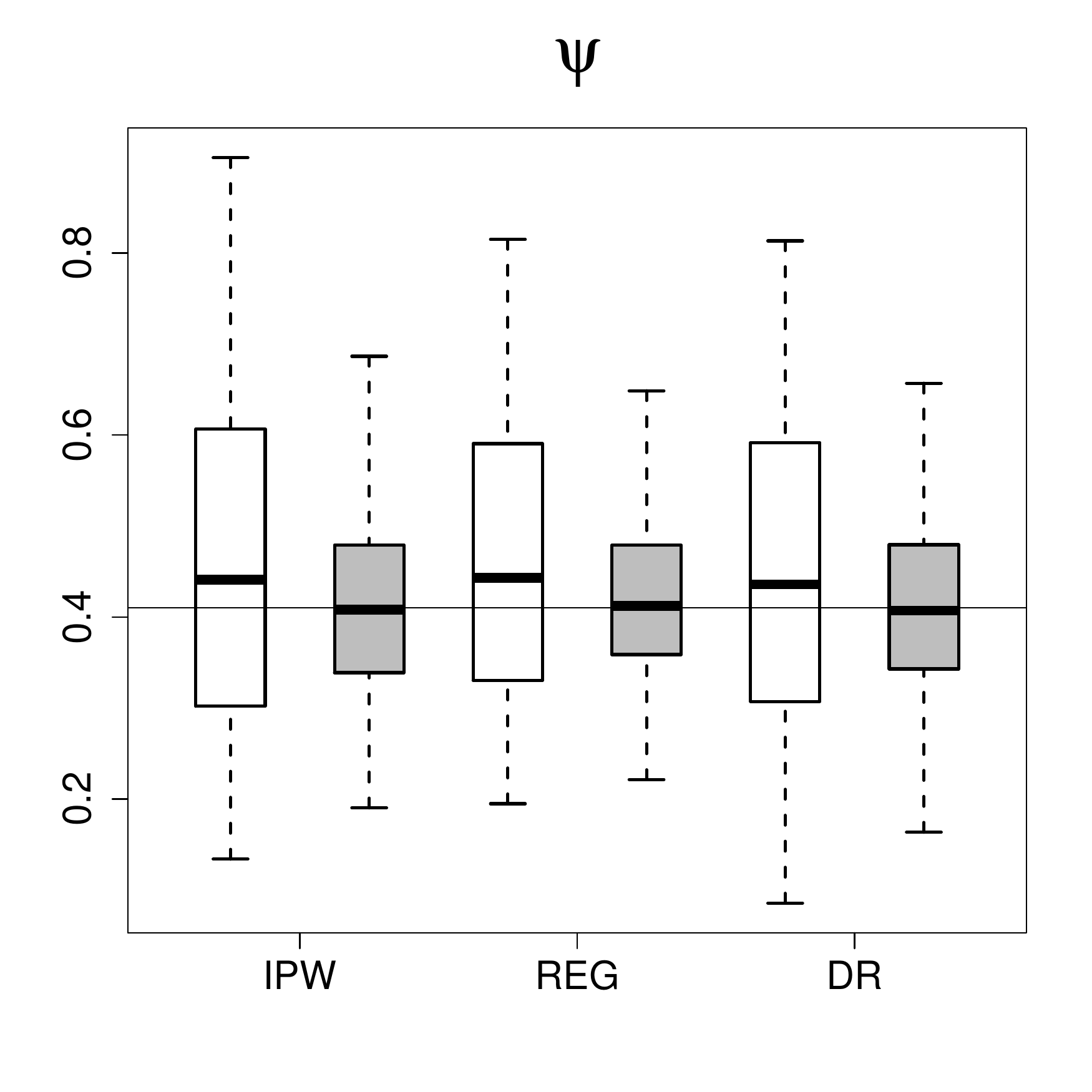}
  \caption{Both outcome regression and extended propensity score are correctly specified}
  \label{pic: bin1_pi_tru_mu_tru}
\end{subfigure}

\begin{subfigure}[b]{.47\textwidth}
\centering\captionsetup{width=.9\linewidth}
  \includegraphics[width=.9\linewidth,height=4.5cm]{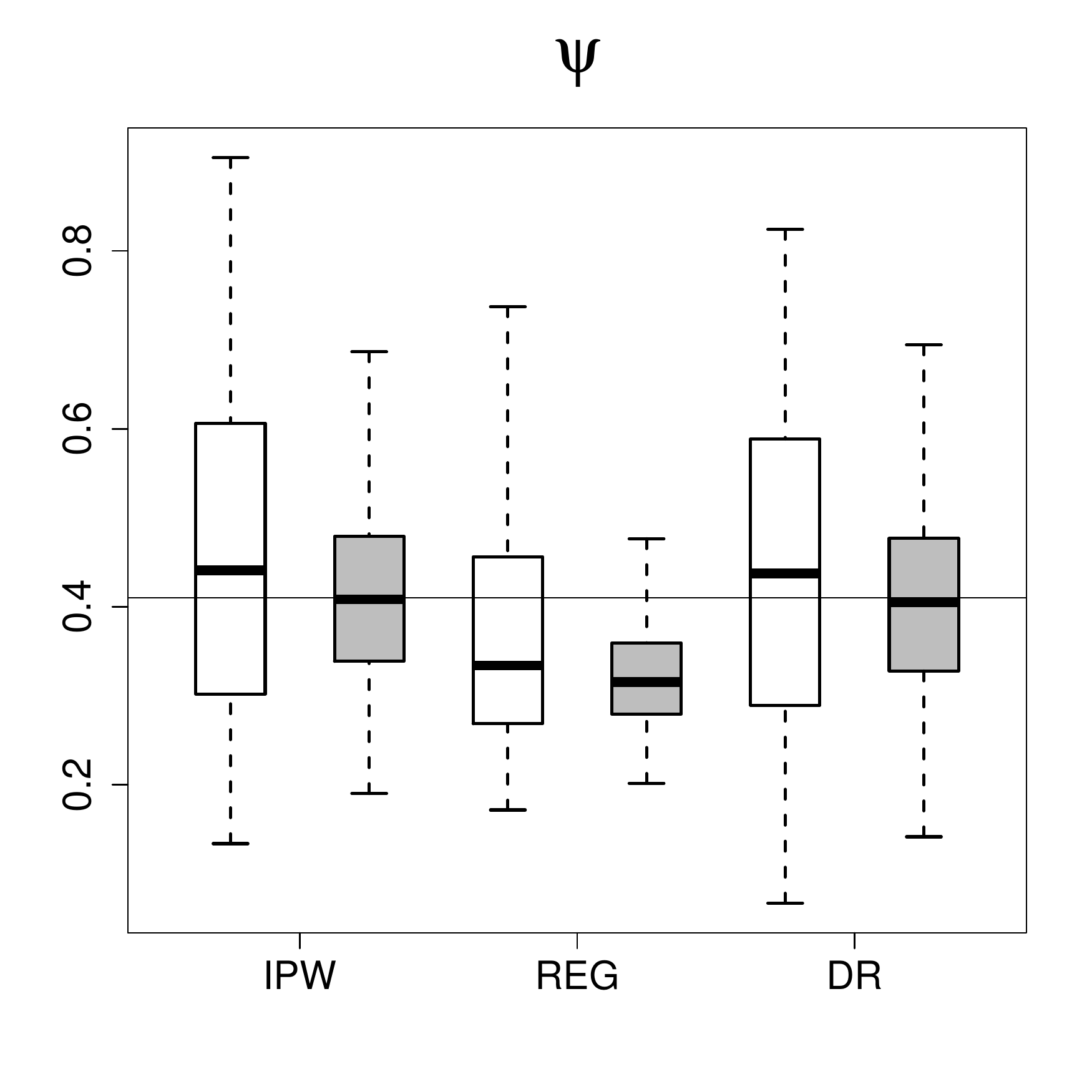}
  \caption{Only the extended propensity score is correctly specified}
  \label{pic: bin1_pi_tru_mu_mis}
\end{subfigure}
\begin{subfigure}[b]{.47\textwidth}
\centering\captionsetup{width=.9\linewidth}
  \includegraphics[width=.9\linewidth,height=4.5cm]{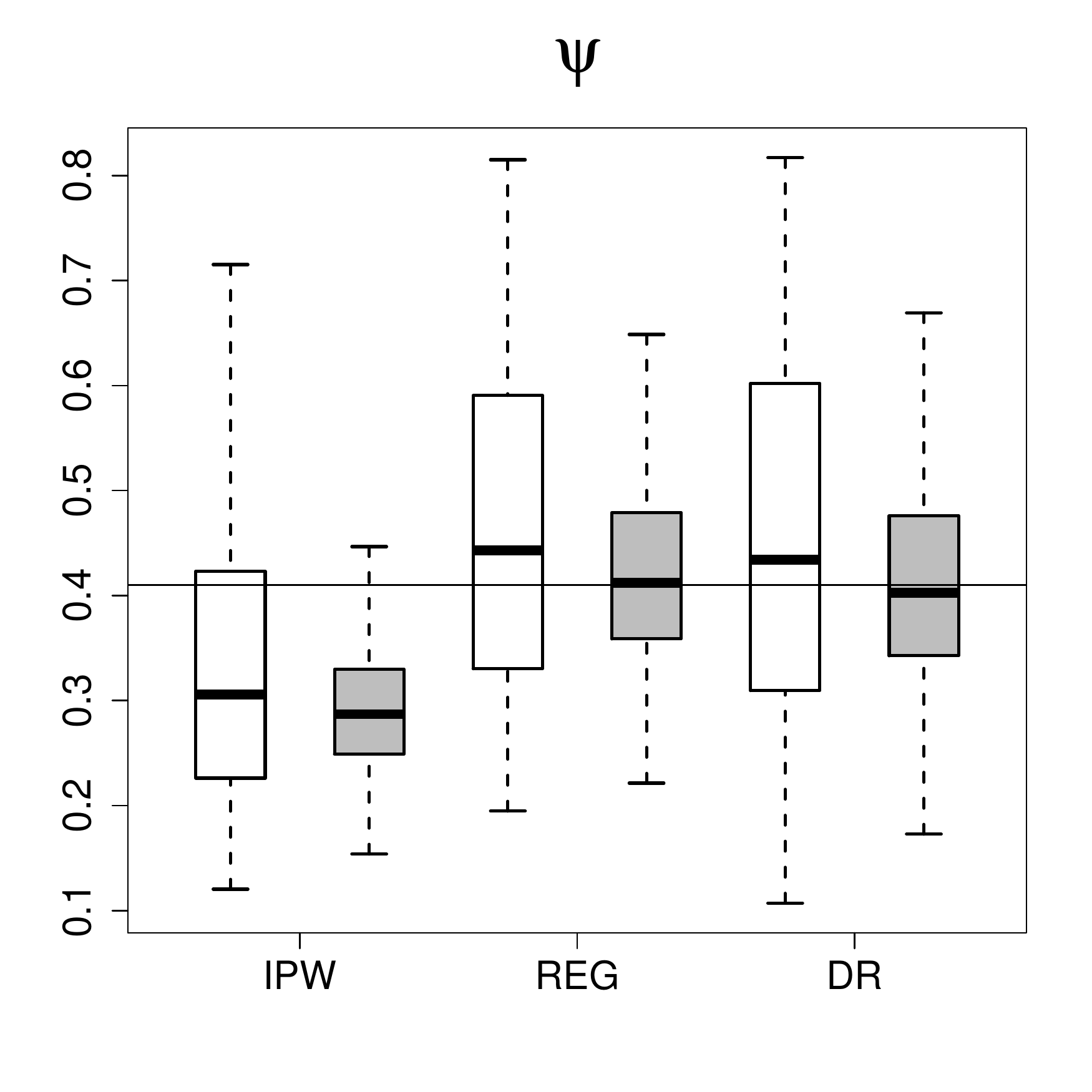}
  \caption{Only the outcome model is correctly specified}
  \label{pic: bin1_pi_mis_mu_tru}
\end{subfigure}%
\caption{Performance of the IPW, OR and DR estimators of $\psi$ with binary outcomes.}
\label{fig: binary_sim}
\floatfoot{Note: In each boxplot, the true value $\psi_0$ is marked by the horizontal lines, white boxes are for $n = 1000$ and grey boxes are for $n = 5000$.}
%\ref{pic: NMAR}: data generated assuming NMAR. Fig
%\ref{pic: MAR}: data generated assuming MAR with $X$ always observed and $Y$  missing at random. For both Fig \ref{pic: NMAR} and \ref{pic: MAR}, We calculated naive estimatior
%$\mu_{naive}$, NMAR estimator $\hat\mu_{NMAR}$, and
%pairwise MAR estimator $\hat\mu_{PMAR}$. In each boxplot, white boxes are for
%sample size 1000, and gray ones for 5000. The horizontal line marks the true value of the
%parameter.}
\end{figure}
% latex table generated in R 3.3.0 by xtable 1.8-2 package
% Sat Jul  2 23:04:54 2016
\begin{table}[ht]
\begin{threeparttable}[c]
\renewcommand{\TPTminimum}{\linewidth}
%\centering
\begin{centering}
\caption{Empirical coverage rates based on 95\% Wald confidence intervals for both binary and continuous outcomes}
\label{tb: CI}
\begin{tabular}{ccccccrrrrrr}
\cmidrule[\heavyrulewidth]{7-12}
&&&&&& & \multicolumn{2}{c}{Binary $Y$} && \multicolumn{2}{c}{Cont. $Y$} \\ 
\cmidrule[\lightrulewidth]{7-12}
&&&&&&  sample size $(n)$ & 1000 & 5000 && 1000 & 5000 \\ 
\cmidrule[\lightrulewidth]{7-12}
&&&&&&\multicolumn{6}{l}{(i) both $\pi$ and $\mu$ are correct}\\
&&&&&&$\hat{\psi}^{ipw}$ & 0.86 & 0.90 && 0.96 & 0.95 \\ 
&&&&&&$\hat{\psi}^{reg}$& 0.84 & 0.92 && 0.97 & 0.95 \\ 
&&&&&&$\hat{\psi}^{DR}$& 0.85 & 0.91 && 0.97 & 0.96 \\ 
&&&&&&\multicolumn{6}{l}{(ii) only $\pi$ is correct}\\
&&&&&&$\hat{\psi}^{ipw}$ & 0.86 & 0.90 && 0.96 & 0.95 \\ 
&&&&&&$\hat{\psi}^{reg}$& 0.79 & 0.60 && 0.39 & 0.00 \\ 
&&&&&&$\hat{\psi}^{DR}$ & 0.86 & 0.91 && 0.97 & 0.95 \\ 
&&&&&&\multicolumn{6}{l}{(iii) only $\mu$ is correct}\\
&&&&&&$\hat{\psi}^{ipw}$& 0.78 & 0.53 && 0.39 & 0.00 \\ 
&&&&&&$\hat{\psi}^{reg}$& 0.84 & 0.92 && 0.97 & 0.95 \\ 
&&&&&&$\hat{\psi}^{DR}$& 0.85 & 0.92 && 0.96 & 0.96 \\ 
\cmidrule[\heavyrulewidth]{7-12}
\end{tabular}
   \end{centering}
\begin{tablenotes}
\item The coverage was evaluated under three scenarios: (i) both outcome regression and the extended propensity score are correctly specified, in (ii) only the extended propensity score is correct and in (iii) only the outcome regression model is correct.
\end{tablenotes}
\end{threeparttable}
\end{table}
%\footnote{xx}

The data generating mechanism described in Step 1 satisfies the assumptions (IV.1)--(IV.2) for both $a=0,1$. As shown in example \ref{eg: saturate_logit}, $\psi$ is identified from the observed data since the treatment mechanism is a separable logit model. Also in the Supplementary Materials, we verify that model \eqref{eq: Y0_correct} for $E(Y|A=0,C,Z)$ contains the true data generating mechanism. Simulations for scenario (ii) were similar to scenario (i) except that \eqref{eq: propen_correct} was replaced with

\begin{equation}\label{eq: propen_mis}
\logit \Pr(A=1|Y_0,Z,C;\gamma)=\theta_1+\theta_2Z+\theta_3C_1+\eta Y_0,
\end{equation}
which is misspecified if $\theta_4\neq0$ in equation \eqref{eq: propen_correct}. For scenario (iii), the potential outcome model \eqref{eq: Y0_correct} was replaced with
\begin{equation}\label{eq: Y0_mis}
\logit E(Y|A=0,Z,C;\xi)=\xi_1+\xi_2C_1+\xi_4Z,
\end{equation}
which is misspecified if $\xi_3\neq0$ and $\xi_5\neq0$ in equation \eqref{eq: Y0_correct}. We use the R package BB \citep{varadhan2009bb} to solve the nonlinear estimating equations. Simulation results for 1000 Monte Carlo samples are reported in Figure \ref{fig: binary_sim} and empirical coverage rates are presented in Table \ref{tb: CI}. Under correct model specification, all estimators have negligible bias which diminishes with increasing sample size. In agreement with our theoretical results, the IPW and regression estimators are biased with poor empirical coverages when the extended propensity score or the outcome model is mis-specified, respectively. The DR estimator performs well in terms of bias and coverage when either model is mis-specified but the other is correct. When all models are correctly specified, the relative efficiency of the locally semiparametric  efficient estimator compared to the DR estimator of $\eta$ and $\psi$ are 0.840 and 0.810 respectively, based on Monte Carlo standard errors at sample size $n=5000$. This shows that substantial efficiency gain may be possible at the intersection submodel when using the locally efficient score.

%At the sample size of $n=5000$, when only the propensity model was misspecified, the IPW estimator had a bias of $0.116$ while the outcome regression estimator had negligible bias equal to $0.010$. When only the outcome regression model was misspecified, the regression estimator had a bias of $0.085$ while the IPW estimator had negligible bias equal to $0.003$. The DR estimator provides consistent results when either model was correct with biases of $5e^{-4}$ and $0.006$ respectively in the above situation and $0.001$ when both models were correctly specified. {\blue The relative efficiency for selection bias parameter is 0.840 comparing the efficient estimator with DR and for $\psi$ is 0.810 if $Z|C$ is saturated. When the $Z|C$ is not saturated, the relative efficiency is 0.844 for the selection bias parameter and 0.812 for $\psi$.}
 
% More specifically, MCSE is the empirical standard error among the corresponding estimators obtained from 1000 samples and ASE is the average of 1000 variance estimators using M-estimator method. 

Simulations for a continuous outcome were conducted similarly as for the binary outcome in the following steps. 
\begin{enumerate}
\item[Step $1^{*}$:] Covariates $C_1$ and $C_2$ were generated as in Step 1, $Z$ was generated from model $\logit \Pr(Z=1|C)=0.7+0.8C_1-C_2$, and $Y_0$, $Y_1$ from models $Y_0|Z,C\sim N(0.5+C_1+3C_2,1)$ and $Y_1|Z,C\sim N(1.1-1.3C_1,1)$, $A$ was generated from $\logit \Pr(A=1|Y_0,Z,C)=-0.2-3Z-3C_1+0.3Y_0+4C_1Z$, and $Y=Y_0(1-A)+Y_1A$.

\item[Step $2^{*}$:] Same as Step 2.

\item[Step $3^{*}$:] Same as Step 3 except the following regression outcome models were fit to the data.
\begin{eqnarray}
E\{Y\exp(\eta Y)|A=0,Z,C;\xi\}=&\xi_1+\xi_2C_1+\xi_3C_2+\xi_4Z+\xi_5C_1Z\nonumber\\
&+\xi_6C_2Z+\xi_7C_1C_2+\xi_8C_1C_2Z.\label{eq: Y0_correct_cont_YexpY}
\end{eqnarray}

%\vspace{-10mm}
\begin{eqnarray}
E\{\exp(\eta Y)|A=0,Z,C;\xi\}=\xi_9+\xi_{10}C_1+\xi_{11}C_2+\xi_{12}Z+\xi_{13}C_1Z\nonumber\\
+\xi_{14}C_2Z+\xi_{15}C_1C_2+\xi_{16}C_1C_2Z.\label{eq: Y0_correct_cont_expY}
\end{eqnarray}

\item[Step $4^{*}$:] Same as Step 4 except that \eqref{eq: Y0_correct} was replaced by \eqref{eq: Y0_correct_cont_YexpY} and \eqref{eq: Y0_correct_cont_expY}.

\item[Step $5^{*}$:] Same as Step 5.

\end{enumerate}

\begin{figure}
\centering
\begin{subfigure}[b]{.47\textwidth}
%  \centering
  \includegraphics[width=.9\linewidth,height=4.5cm]{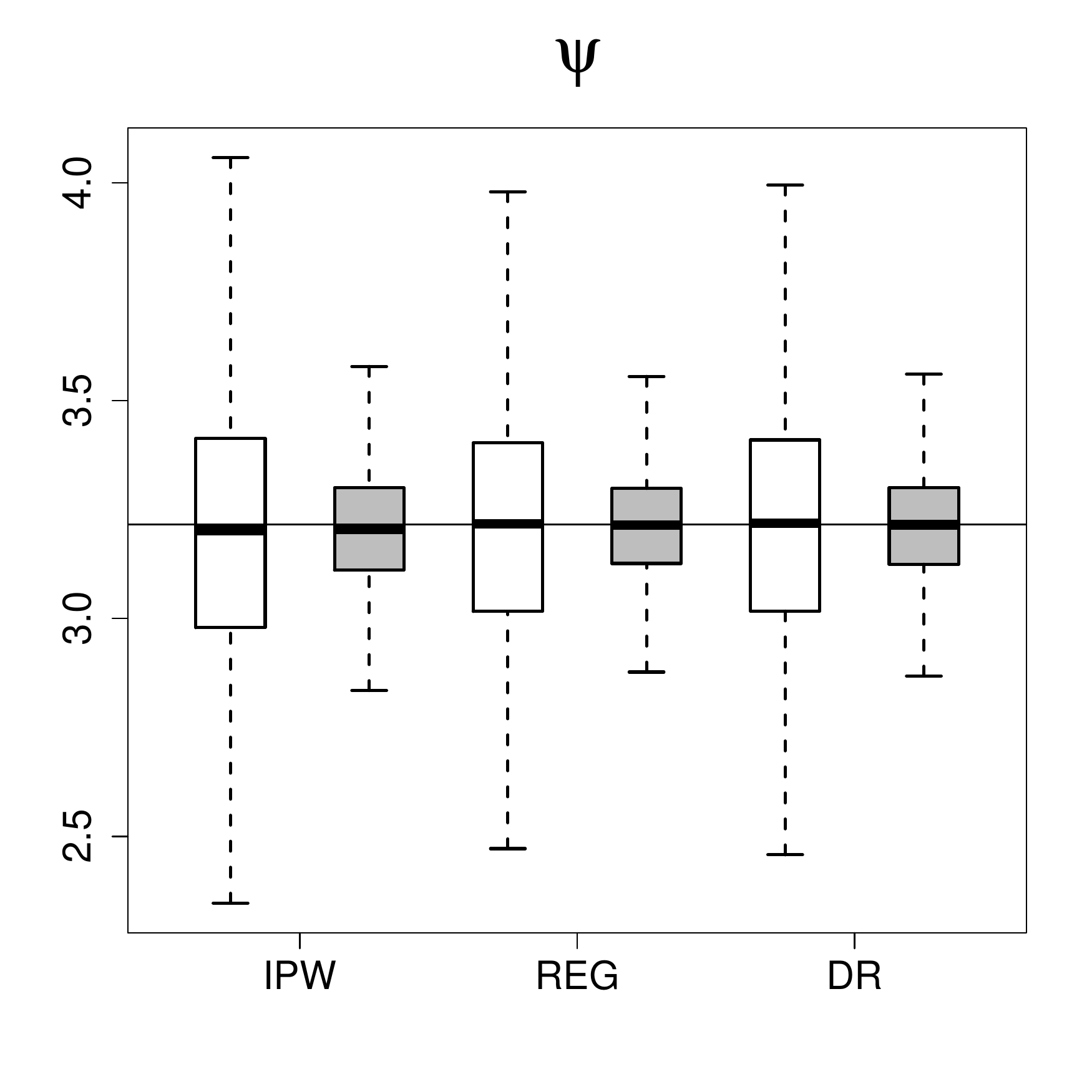}
  \caption{Both models are correctly specified}
  \label{pic: bin0_pi_tru_mu_tru}
\end{subfigure}

\begin{subfigure}[b]{.47\textwidth}
\centering\captionsetup{width=.9\linewidth}
  \includegraphics[width=.9\linewidth,height=4.5cm]{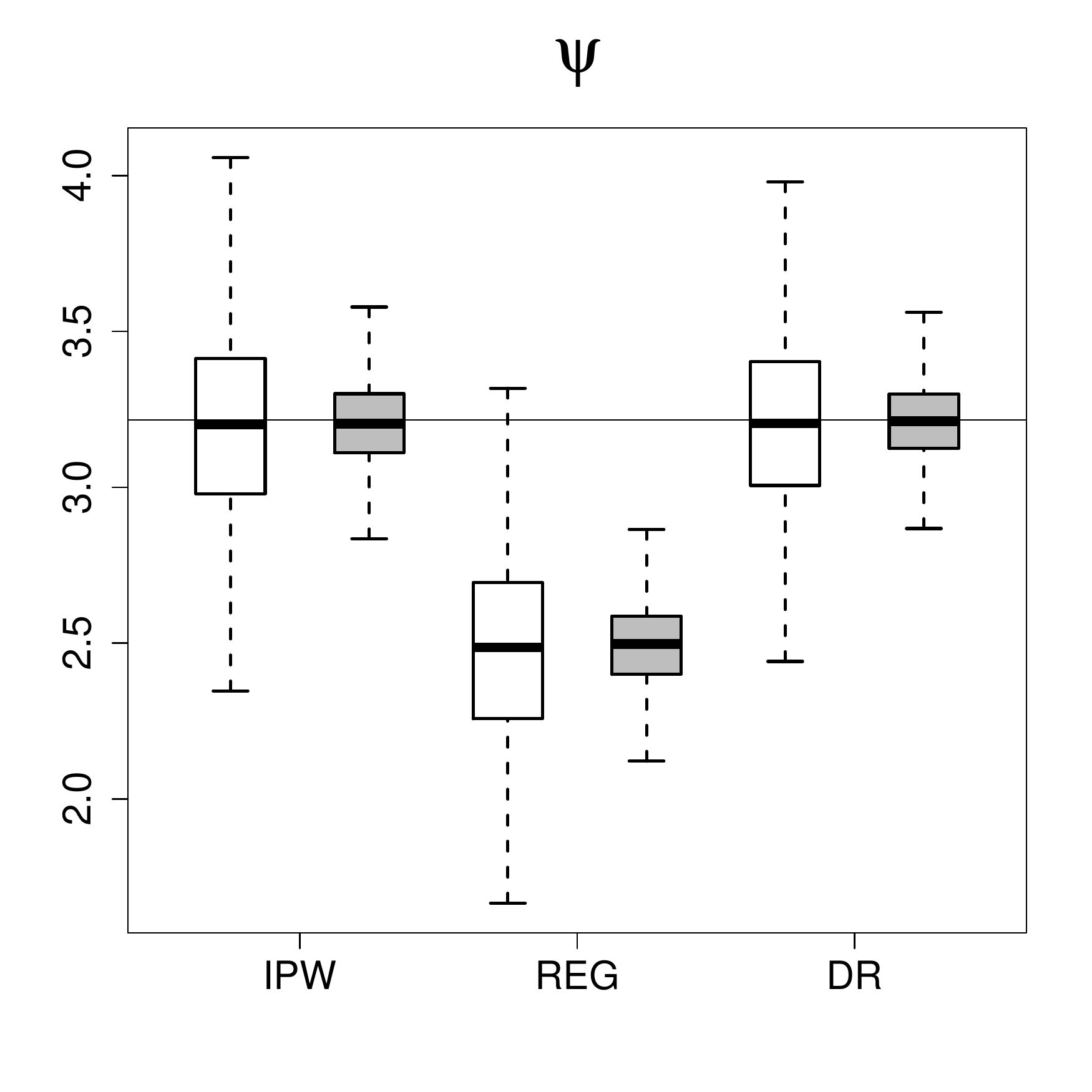}
  \caption{Only the extended propensity score is correctly specified}
  \label{pic: bin0_pi_tru_mu_mis}
\end{subfigure}
\begin{subfigure}[b]{.47\textwidth}
\centering\captionsetup{width=.9\linewidth}
  \includegraphics[width=.9\linewidth,height=4.5cm]{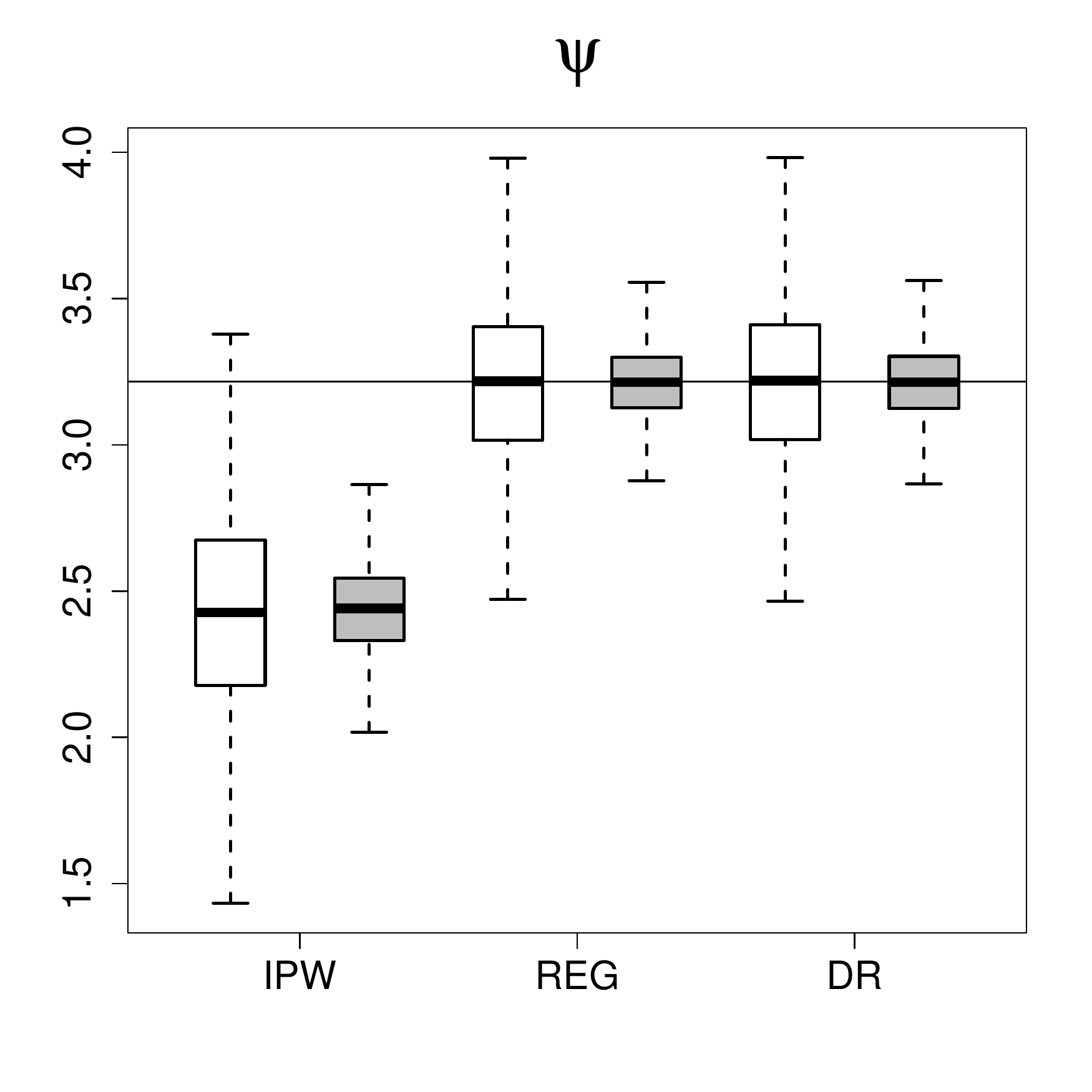}
  \caption{Only the outcome model is correctly specified}
  \label{pic: bin0_pi_mis_mu_tru}
\end{subfigure}%
\caption{Performance of the IPW, OR and DR estimators of $\psi$ with continuous outcomes}
\label{fig: cont_sim}
\floatfoot{Note: In each boxplot, the true value $\psi_0$ is marked by the horizontal lines, white boxes are for $n = 1000$ and grey boxes are for $n = 5000$.}
%\floatfoot{Fig
%\ref{pic: NMAR}: data generated assuming NMAR. Fig
%\ref{pic: MAR}: data generated assuming MAR with $X$ always observed and $Y$  missing at random. For both Fig \ref{pic: NMAR} and \ref{pic: MAR}, We calculated naive estimatior
%$\mu_{naive}$, NMAR estimator $\hat\mu_{NMAR}$, and
%pairwise MAR estimator $\hat\mu_{PMAR}$. In each boxplot, white boxes are for
%sample size 1000, and gray ones for 5000. The horizontal line marks the true value of the
%parameter.}
\end{figure}

%%\vspace{-4mm}
Simulation for a continuous outcome under scenario (ii) was carried out similarly as that for scenario (i) except that \eqref{eq: propen_correct} was replaced by \eqref{eq: propen_mis}. For scenario (iii), the potential outcome models \eqref{eq: Y0_correct_cont_YexpY} and \eqref{eq: Y0_correct_cont_expY} were replaced with the linear models
\begin{equation}\label{eq: Y0_mis_cont_YexpY}
E\{Y\exp(\eta Y)|A=0,Z,C;\xi\}=\xi_1+\xi_2C_1+\xi_4Z.
\end{equation}

%\vspace{-6mm}
\begin{equation}\label{eq: Y0_mis_cont_expY}
E\{\exp(\eta Y)|A=0,Z,C;\xi\}=\xi_9+\xi_{10}C_1+\xi_{12}Z.
\end{equation}
We use the R package nleqslv \citep{nleqslv2014} to solve the nonlinear estimating equations.

We verify in the Example \ref{eg: sep_binIV_contY} of the Supplementary Materials that $\psi$ is identified from the observed data. The simulation results for 1000 Monte Carlo samples are reported in Figure \ref{fig: cont_sim} and empirical coverage rates are presented in Table \ref{tb: CI}. Results are similar to the results for the binary outcome. Under correct model specification, all estimators have negligible bias which diminishes with increasing sample size. The IPW and OR estimators are biased with poor empirical coverages when the corresponding model is mis-specified. The DR estimator performs well in terms of bias and coverage when either the extended propensity score or the outcome regression model is correctly specified.

%The results of 1000 Monte Carlo simulated samples are provided in Table \ref{tb: ipw_reg_DR_consistency_cont_y_apr14th}. At sample size $n=1000$, when only the propensity score model was misspecified, the IPW estimator suffered from a bias of $0.806$, while the outcome regression estimator had negligible bias equal to $0.009$. When the outcome regression model was misspecified, the regression estimator had a bias of $0.738$ while the IPW estimator had negligible bias equal to $0.024$. The DR estimator was consistent in the union model with bias comparable to the consistent estimator under a given scenario.
% Note that when both models were wrong, the DR still had relatively smaller bias in the simulations for both Table \ref{tb: ipw_reg_DR_consistency} and \ref{tb: ipw_reg_DR_consistency_cont_y_apr14th} with bias 0.029 and 0.010 respectively for each of these scenarios at sample size $n=1000$. This may not be the case for other data generating mechanisms. For sample size $n<1000$, the solver in the R packages used in the simulations did not always converge, for a summary of convergence rates see the Supplementary Materials. However, this was less of an issue for moderate to large sample sizes.

%\vspace{-4mm}
\section{\bf Application}\label{sec: application}
Since the 1980s, tax-deferred programs such as individual Retirement Accounts (IRAs) and the 401(k) plan have played an important role as a channel for personal savings in the United States. Aiming to encourage investment for future retirement, the 401(k) plan offers tax deductions on deposits into retirement accounts and tax-free accrual of interest. The 401(k) plan shares similarities with IRAs in that both are deferred compensation plans for wage earners but the 401(k) plan is only provided by employers. The study includes 9275 people and once offered the 401(k) plan, individuals decide whether to participate in the program. However, participants usually have a stronger preference for savings which suggests the presence of selection bias. This was addressed as individual heterogeneity by \cite{abadie2003semiparametric} and it has been pointed out that a simple comparison of personal savings between participants and non-participants may yield results that were biased upward. It was also postulated that given income, the 401(k) eligibility is unrelated to the individual preferences for savings thus can be used as an instrument for participation in 401(k) program \citep{poterba1994401,poterba1995401}. The complier causal effect for the 401(k) plan was studied by \cite{abadie2003semiparametric}. Here, we reanalyze these data to illustrate the proposed estimators of the marginal ETT.

We illustrate the methods in the context of a dichotomous outcome defined as the indicator that a person falls in the first quartile of net savings of the observed sample (equal to $-\$500$). 
The treatment variable is a binary indicator of participation in a 401(k) plan and the IV is a binary indicator of 401(k) eligibility. The covariates are standardized log family income ($\log_{10}\text{(income)}-4.5$), standardized age ($\text{age}-41$) and its square, marital status and family size. Age ranged from 25 to 64 years, marital status is binary indicator variable and family size ranges from 1 to 13 people. These covariates are thought to be associated with unobserved preferences for savings. Let $\psi=E(Y_0|A=1)$ denote for a family that actually participated in the 401(k) program, the probability that they would have had net financial assets above the first quartile, had possibly contrary to fact, they been forced not to participate in the program. The $\text{ETT}=E(Y_1-Y_0|A=1)$ is the effect of 401(k) plan on the difference scale for the probability of family net financial assets above the first quartile among participants. Equivalently, ETT can also be interpreted as an effect of the intervention in reducing a person's risk for poor savings performance as measured by falling below the first quartile of the empirical distribution of savings for the sample. Before implementing our IV estimators, we first obtained a standard IPW estimator of the ETT under an assumption of no unmeasured confounding, i.e. $\hat\psi^{ipw}_0$ defined as $\hat\psi^{ipw}$ with $\alpha=0$. Thus, the extended propensity score was modeled as:
\[\logit \Pr(A=1|Z,C)=1+Z+\log(\text{income})+\text{married}+\text{age}+\text{fsize}+\text{age}^2,\]
and estimated by standard maximum likelihood. The IPW estimate of $\psi$ was $\hat{\psi}^{ipw}_0=0.688$ with standard error ({se}) $0.014$, where se was evaluated using the sandwich estimator accounting for all sources of variabilities. In comparison, the estimator based on the empirical estimate of $E(Y|A=1)$ was 0.883 ($\text{se}=0.006$). 
Thus an estimate of ETT was $\widehat{\text{ETT}}=0.194$ ($\text{se}=0.016$), which suggests the 401(k) plan may have a significant effect on increasing the family net financial assets among participants.

However, this result may be spurious due to the suspicion that even after controlling for observed covariates, there may still exist unmeasured factors that confound the relationship between 401(k) plan and the family net financial assets. Assuming assumptions (IV.1)--(IV.2) and condition 1, we applied the methods proposed in Section \ref{sec: inferences} to estimate the ETT in the presence of unmeasured confounders. The following parametric models were considered:
\[\logit \Pr(Z=1|C)=1+\log(\text{income})+\text{married}+\text{age}+\text{fsize}+\text{age}^2,\] 
\[\logit \Pr(Y=1|A=0,Z,C)=1+Z+\log(\text{income})+\text{married}+\text{age}+\text{fsize}+\text{age}^2,\]
We specified the selection bias function as in \eqref{eq: alpha=Y0}, thus the selection bias function was assumed to depend on $Y_0$ linearly. Possible deviations from this simple model was explored by allowing for potential interactions of $Y_0$ with observed covariates in the extended propensity score. Thus, we posited the following parametric model for the extended propensity score which satisfies identifying condition \ref{cond: not1} as a submodel of the separable model:
\[\logit \Pr(A=1|Y_0,Z,C)=1+Z+Y_0+\log(\text{income})+\text{married}+\text{age}+\text{fsize}+\text{age}^2,\]
Table \ref{tb: retirement} reports point estimates and estimated standard errors for the IV, extended propensity score and the outcome regression models. Although the DR estimator also involves an outcome regression model among the unexposed, it is the same model as required for the regression estimator, thus these estimates are only repeated once. The instrument is strongly associated with family income ($\log\text{OR}= 2.823 \text{, se}=0.106$), age ($\log\text{OR}=0.007\text{, se}=0.002$) and age square ($\log\text{OR}=-0.002\text{, se}=2e^{-4}$). The selection bias parameter was estimated to be 0.320 ($\text{se}=0.115$) by IPW, 0.385 ($\text{se}=0.135$) by OR and 0.280 ($\text{se}=0.101$) by DR estimation. This provides strong evidence that unmeasured confounding may be present and the stronger saving preference one has, the more likely one would participate in the 401(k) plan. All three estimators of the marginal ETT also agree with each other: they are significant but with a smaller Z-score value than when the selection bias is ignored (for example, the IPW estimator suggests $\widehat{\text{ETT}}=0.134$, $\text{se}=0.013$). The efficient estimator for the selection bias parameter is 0.273 and for the ETT is 0.137, both in agreement with the other three estimators. Thus we may conclude that even after adjustment for unobserved preferences for savings, the 401(k) plan still can increase net financial assets among participants.

These findings roughly agree with results obtained by Abadie in the sense that the IV estimate corrects the observational estimate towards the null. However, it may be difficult to directly compare our findings to those of Abadie who reported the compliers average treatment effect under a monotonicity assumption of the IV-exposure relationship, and assuming no unmeasured confounding of this first stage relation. Our approaches rely on neither assumption, but instead rely on condition \ref{cond: not1} encoded in the functional form of the extended propensity score model for identification. In order to assess the robustness of the selection bias model, additional functional forms were explored. We considered adding to $\alpha$ an interaction between $Y_0$ and each of the covariates: log income, marriage status, family size. There was no evidence in favor of any such interaction. 
\begin{table}[h]
\centering
\caption{Point estimates and estimated se [in bracket] of IPW, OR and DR estimators for ETT of 401(k) plan as well as the parameters for IV, extended propensity score and outcome regression outcome models required by those estimators}
\label{tb: retirement}
\begin{tabular}{lrrrr}
\hline
& IV model & IPW propensity & regression & DR propensity \\
\hline
Intercept & -0.180 [0.058] & -8.685 [1.832] & 1.307 [0.073] & -8.629 [1.796] \\ 
linc & 2.695 [0.107] & 1.626 [0.210] & 0.618 [0.128] & 1.633 [0.209] \\ 
age & 0.007 [0.002] & -0.009 [0.005] & 0.035 [0.003] & -0.009 [0.005] \\ 
fsize & -0.037 [0.019] & -0.004 [0.033] & -0.127 [0.022] & -0.005 [0.033] \\ 
marr & -0.145 [0.063] & -0.032 [0.108] & -0.133 [0.075] & -0.031 [0.108] \\ 
$\text{age}^2$ & -0.002 [2e-04] & 0.001 [4e-04] & 6e-04 [3e-04] & 0.001 [4e-04] \\ 
Z & & 9.150 [1.820] & -0.210 [0.074] & 9.126 [1.781] \\ 
$\alpha$ & & 0.320 [0.115] & 0.385 [0.135] & 0.280 [0.101] \\ 
\hline
\multicolumn{2}{l}{$\psi=E(Y_0|A=1)$} & 0.749 [0.012] & 0.746 [0.012] & 0.750 [0.012] \\ 
ETT & & 0.134 [0.013] & 0.137 [0.014] & 0.132 [0.014] \\ 
%ETT & & 0.134 [0.014] & 0.137 [0.014] & 0.132 [0.014] \\ 
\hline
\end{tabular}
\end{table}

%\vspace{-3mm}
\section{\bf Discussion}\label{sec: dis}
In this paper, we establish that access to an IV allows for identification of an association between exposure to the treatment and the potential outcome when unexposed, which directly encodes the magnitude of selection bias into treatment due to confounding. We propose IPW, OR as well as DR estimators for the treatment effect amongst treated individuals. \cite{vansteelandt2003causal} and \cite{robins1994correcting} proposed identification and inference approaches under no-current treatment value interaction assumption, thus their estimators remain consistent under the null hypothesis of no ETT. In contrast, the identification and inference approaches we proposed may be particularly valuable when an ITT analysis indicates a non-null treatment effect and thus Robins' identification assumption of no-current treatment value interaction may be violated.

The proposed methods assume the treatment is binary. They can be generalized without much effort to categorical treatment. However, when the treatment is continuous (for example, $A$ is treatment dose), then a parametric model for the treatment effect as well as a model for the density of $A$ may be unavoidable for estimation. We leave this as a topic for future research.

%\section*{Acknowledgements}
%And this is an acknowledgements section with a heading that was produced by the

%\vspace{-5mm}

\bibliographystyle{imsart-nameyear}

%\bibliographystyle{apa}
%\bibliography{/Users/Lan/Desktop/LAN/study/statistics/biostatistics/research/bibliography/mybib}
\bibliography{/Users/Lan/Sync/Lan/study/statistics/biostatistics/research/bibliography/mybib}

\newpage
\appendix

Appendix \ref{sec: appendix_prop_proof} contains proofs of the propositions.
Appendix \ref{sec: appendix_identi} presents proofs of the examples in the main text, and more examples about identification of the models.
Appendix \ref{sec: appendix_more_deri} presents more derivations mentioned in the main text.
Appendix \ref{sec: appendix_efficiency} presents derivations of semiparametric efficiency theory.

\section{\bf Proofs of propositions}\label{sec: appendix_prop_proof}

\vspace{3mm}\noindent\textbf{Proof of Proposition \ref{prop: indenti}}

%\begin{proof}
We prove by contradiction. Suppose we have two candidates $\Pr_1(A,Y_0,Z,C)$ and $\Pr_2(A,Y_0,Z,C)$ satisfying the same observed density:
\[{\Pr}_1(A,Y_0,Z,C)={\Pr}_2(A,Y_0,Z,C).\]
By the assumption (IV.2), we have the decomposition for the joint distribution:
\[f_j(A,Y_0,Z,C)=f_j(C)f_j(Z|C)f_j(Y_0|C)f_j(A|Y_0,Z,C) \text{ for } j=1,2.\]
Since $f(C)$ and $f(Z|C)$ can be identified from the observed data, we have $f_1(C)=f_2(C)$ and $f_1(Z|C)=f_2(Z|C)$. Thus,
\[f_1(Y_0|C){\Pr}_1(A=0|Y_0,Z,C)=f_2(Y_0|C){\Pr}_2(A=0|Y_0,Z,C),\]
and equivalently
\[\frac{{\Pr}_1(A=0|Y_0,Z,C)}{{\Pr}_2(A=0|Y_0,Z,C)}=\frac{f_2(Y_0|C)}{f_1(Y_0|C)}.\]
The equation contradicts the condition that we require the ratios unequal. Thus, that the ratios are not equal is equivalent to the impossibility of two sets of candidates satisfying the same observed quantities, i.e. the identifiability of the joint distribution.
%\end{proof}

\noindent\textbf{Proof of Proposition \ref{prop: ipw_unbias}}
%\begin{proof}

We first prove equation \eqref{eq: ipw_identity}. Note that
\begin{eqnarray*}
&& E\biggl\{\frac{{\pi}(Y_0,Z,C)Y_0(1-A)}{\Pr(A=1)(1-{\pi}(Y_0,Z,C))}\biggr\}\\
&=& E\biggl\{\frac{{\pi}(Y_0,Z,C)Y_0}{\Pr(A=1)}\biggr\}\\
&=& E\biggl\{\frac{AY_0}{\Pr(A=1)}\biggr\}\\
&=& E(Y_0|A=1)\\
&=&\psi.
\end{eqnarray*}
Thus, equation \eqref{eq: ipw_identity} is proved.

We show that if $\pi(Y_0,Z,C)$ is correctly specified, the equations \eqref{eq: ipw_intercept}--\eqref{eq: ipw_alpha} hold at the true value $\gamma$ thus they are indeed unbiased estimating equations for $\gamma$. The equality is easy to show for \eqref{eq: ipw_intercept}--\eqref{eq: ipw_C} by the law of iterated expectations. For \eqref{eq: ipw_alpha}, the assumptions (IV.1)--(IV.2) imply $Y_0\independent Z|C$, thus
\begin{eqnarray*}
&&E\bigl[\frac{1-A}{1-\pi(Y_0,Z,C)}t(Y,C)\{l(Z,C)-E(l(Z,C)|C)\}\bigr]\\
&=&E\bigl[\frac{1-A}{1-\pi(Y_0,Z,C)}t(Y_0,C)\{l(Z,C)-E(l(Z,C)|C)\}\bigr]\\
&=&E\bigl[t(Y_0,C)\{l(Z,C)-E(l(Z,C)|C)\}\bigr]\\
&=&E\bigl[E(t(Y_0,C)|C)\{E(l(Z,C)|C)-E(l(Z,C)|C)\}\bigr]\\
&=&0.
\end{eqnarray*}
Thus, by equation \eqref{eq: ipw_identity}, $\hat\psi^{ipw}$ is consistent for $\psi$.

%Note that if the propensity score $\pi(Y_0,Z,C)$ is correctly modeled, the consistency of IPW estimator follows from equation \eqref{eq: ipw_identity}.

Assume 

\begin{equation}\label{eq: cond_uniqueness_ipw}
E\biggl[\frac{\partial }{\partial\gamma^{T}}\frac{1-A}{1-{\pi}(Y_0,Z,C;\gamma)}
\begin{pmatrix}
1 \\
h_1(Z,C)-E(h_1(Z,C)|C) \\
h_2(C)-E(h_2(C))\\
t(Y,C)\{l(Z,C)-E(l(Z,C)|C)\}
\end{pmatrix}\biggr] \text{is invertible.}
\end{equation}

\noindent Condition \eqref{eq: cond_uniqueness_ipw} is sufficient for local uniqueness of nuisance parameter estimates obtained from equations \eqref{eq: ipw_intercept}--\eqref{eq: ipw_alpha} and thus $\psi$ is identified from the observed data.

%\end{proof}

%\newpage

\noindent\textbf{Proof of Proposition \ref{prop: reg_unbias}}
%\begin{proof}

We first prove equation \eqref{eq: moment_eq}. Note that 

%\begin{eqnarray*}
%\exp\{\alpha(Y_0,Z,C)\}&=&\frac{f(Y_0|A=1,Z,C)f(Y_0=0|A=0,Z,C)}{f(Y_0|A=0,Z,C)f(Y_0=0|A=1,Z,C)}\\
%&=&\frac{\Pr(A=1|Y_0,Z,C)\Pr(A=0|Y_0=0,Z,C)}{\Pr(A=0|Y_0,Z,C)\Pr(A=1|Y_0=0,Z,C)}.
%\end{eqnarray*}
%
\vspace{-4mm}
%\noindent Hence,
\begin{eqnarray*}
&&\frac{E[\exp\{\alpha(Y,Z,C)\}g(Y,C)|A=0,Z,C]}{E[\exp\{\alpha(Y,Z,C)\}|A=0,Z,C]}\\
&=&\frac{E[\exp\{\alpha(Y_0,Z,C)\}g(Y_0,C)|A=0,Z,C]}{E[\exp\{\alpha(Y_0,Z,C)\}|A=0,Z,C]}\\
&=&E\Bigl[\frac{f(Y_0|A=1,Z,C)f(Y_0=0|A=0,Z,C)}{f(Y_0|A=0,Z,C)f(Y_0=0|A=1,Z,C)}g(Y,C)|A=0,Z,C\Bigr]\big/\\
&&E\Bigl[\frac{f(Y_0|A=1,Z,C)f(Y_0=0|A=0,Z,C)}{f(Y_0|A=0,Z,C)f(Y_0=0|A=1,Z,C)}|A=0,Z,C\Bigr]\\
&=&E\Bigl[\frac{f(Y_0|A=1,Z,C)}{f(Y_0|A=0,Z,C)}g(Y,C)|A=0,Z,C\Bigr]\big/E\Bigl[\frac{f(Y_0|A=1,Z,C)}{f(Y_0|A=0,Z,C)}|A=0,Z,C\Bigr]\\
&=&E(g(Y_0,C)|A=1,Z,C)/1\\
&=&E(g(Y_0,C)|A=1,Z,C).
\end{eqnarray*}

We then show that equation \eqref{eq: reg_alpha} holds at the true value of $\xi$ and $\eta$ and thus are indeed unbiased estimating equation for $\eta$. Note that by (IV.1)--(IV.2), we have $Y_0\independent Z|C$, thus 
\begin{eqnarray*}
&& E\bigl[\bigl\{w(Z,C)-E(w(Z,C)|C)\bigr\}\bigl\{AE(g(Y_0,C)|A=1,Z,C)+(1-A)g(Y,C)\bigr\}\bigr]\\
&=&E\bigl[\bigl\{w(Z,C)-E(w(Z,C)|C)\bigr\}\bigl\{Ag(Y_0,C)+(1-A)g(Y_0,C)\bigr\}\bigr]\\
&=&E\bigl[\bigl\{w(Z,C)-E(w(Z,C)|C)\bigr\}g(Y_0,C)\bigr]\\
&=&E\bigl[\bigl\{E(w(Z,C)|C)-E(w(Z,C)|C)\bigr\}E(g(Y_0,C)|C)\bigr]\\
&=&0.
\end{eqnarray*} 
Consistency of the regression estimator $\hat{\psi}^{reg}$ follows from equation \eqref{eq: reg_alpha}.

Assume
\begin{equation}\label{eq: cond_uniqueness_reg}
E\{\{\omega(Z,C)-E(\omega(Z,C)|C)\}A\frac{\partial }{\partial\eta}\frac{E(\exp\{\alpha(Y,Z,C;\eta)\}g(Y,C)|A=0,Z,C)}{E(\exp\{\alpha(Y,Z,C;\eta)\}|A=0,Z,C)}\}\text{ is invertible.}
\end{equation}

\noindent Condition \eqref{eq: cond_uniqueness_reg} is sufficient for local uniqueness of an estimator for $\eta$ obtained from equation \eqref{eq: reg_alpha}. 
To see the relationship between \eqref{eq: cond_uniqueness_reg} and the first order derivative of \eqref{eq: reg_alpha}, note that
\begin{eqnarray*}
&&\frac{\partial}{\partial\eta}E[\{\omega(Z,C)-E(\omega(Z,C)|C)\}\{AE(g(Y_0,C)|A=1,Z,C;\eta)+(1-A)g(Y,C)\}]\\
&=&E[\{\omega(Z,C)-E(\omega(Z,C)|C)\}\{A\frac{\partial E(g(Y_0,C)|A=1,Z,C;\eta)}{\partial\eta}+(1-A)g(Y,C)\}]\\
&=&E[\{\omega(Z,C)-E(\omega(Z,C)|C)\}\{A\frac{\partial }{\partial\eta}\frac{E(\exp\{\alpha(Y,Z,C;\eta)\}g(Y,C)|A=0,Z,C)}{E(\exp\{\alpha(Y,Z,C;\eta)\}|A=0,Z,C)}\}].
%&=&E[\{\omega(Z,C)-E(\omega(Z,C)|C)\}\{\frac{A}{E^2(\exp\{\alpha(Y,C)+\tau(Y,Z,C)\}Y|A=0,Z,C)}\\
%&&\frac{\partial }{\partial\eta}\frac{}{E(\exp\{\alpha(Y,C)+\tau(Y,Z,C)\}|A=0,Z,C)}\}]\\
\end{eqnarray*} 
%Thus, assuming \eqref{eq: cond_uniqueness_reg}, $\psi$ is locally identified from the observed data.
%\end{proof}

\noindent\textbf{Proof of Proposition \ref{prop: DR_unknown_alpha0} }
%\begin{proof}

We use the superscript $\ast$ to denote a misspecified model. Otherwise, an expectation or a model is always evaluated at the true value of parameters. Note that by (IV.1)--(IV.2), we have $Y_0\independent Z|C$. If parametric models lie in $\mathcal{M}_a$, then $\tilde \gamma\xrightarrow{p}\gamma$ and 

%\vspace{-15mm}
\begin{eqnarray*}
&&E\biggl[\bigl\{\omega(Z,C)-E(\omega(Z,C)|C)\bigr\}\tilde Q_g(Y,A,Z,C;\tilde\gamma,\hat\xi)\biggr]\\
&\xrightarrow{p}&E[\bigl\{\omega(Z,C)-E(\omega(Z,C)|C)\bigr\}g(Y_0,C)]\\
&=&E[\bigl\{E(\omega(Z,C)|C)-E(\omega(Z,C)|C)\bigr\}g(Y_0,C)]=0.
%&=&0.
\end{eqnarray*}

%\vspace{-3mm}
\noindent Additionally,

%\vspace{-13mm}
\begin{eqnarray*}
&&\hat{\psi}^{DR}\\
&\xrightarrow{p}&E\biggl\{\frac{\pi(Y_0,Z,C)\{Y_0-E^{\ast}(Y_0|A=1,Z,C)\}}{\Pr(A=1)}+\frac{E^{\ast}(Y_0|A=1,Z,C)\pi(Y_0,Z,C)}{\Pr(A=1)}\biggr\}\\
&=&E(\frac{\pi(Y_0,Z,C)Y_0}{\Pr(A=1)})\\
&=&E(\frac{AY_0}{\Pr(A=1)})\\
&=&E(Y_0|A=1)=\psi.
\end{eqnarray*}

\noindent Thus, $\hat\eta^{DR}$ and $\hat\psi^{DR}$ are consistent if the parametric models lie in $\mathcal{M}_a$.

%\vspace{-2mm}
If parametric models lie in $\mathcal{M}_y$: 
\begin{eqnarray*}
&&E\biggl[\bigl\{\omega(Z,C)-E(\omega(Z,C)|C)\bigr\} \tilde Q_g(Y,A,Z,C;\tilde\gamma,\hat\xi)\biggr]\\
&\xrightarrow{p}&
%E\biggl[\bigl(\omega(Z,C)-E(\omega(Z,C)|C)\bigr)\bigl\{(1-A)e^{\alpha(Y_0,Z,C)+\beta^{\ast}(Z,C)}(g(Y_0,C)-E(g(Y_0,C)|A=1,Z,C))+\\
%&&\quad\quad AE(g(Y_0,C)|A=1,Z,C)+(1-A)g(Y_0,C)\bigr\}\biggr]\\
%&=&
E\biggl[\bigl\{\omega(Z,C)-E(\omega(Z,C)|C)\bigr\}\\
&&\Bigl\{(1-A)\exp\{\alpha(Y_0,Z,C)+\beta^{\ast}(Z,C)\}\{g(Y_0,C)-\frac{E(g(Y_0,C)\exp({\alpha}(Y_0,Z,C))|A=0,Z,C)}{E(\exp({\alpha}(Y_0,Z,C))|A=0,Z,C)}\}\\
&&+ AE(g(Y_0,C)|A=1,Z,C)+(1-A)g(Y_0,C)\Bigr\}\biggr]\\
%&=&E\biggl[\bigl(\omega(Z,C)-E(\omega(Z,C)|C)\bigr)\biggl\{(1-A)e^{\beta^{\ast}(Z,C)}\biggl(e^{\alpha(Y_0,C)}Y_0-e^{\alpha(Y_0,C)}\frac{E(Y_0e^{\hat{\alpha}(Y_0,C)}|A=0,Z,C)}{E(e^{\hat{\alpha}(Y_0,C)}|A=0,Z,C)}\biggr)+\\
%&&\quad\quad AE(Y_0|A=1,Z,C)+(1-A)Y_0\biggr\}\biggr]\\
&=&E\biggl[\bigl\{\omega(Z,C)-E(\omega(Z,C)|C)\bigr\}\biggl\{AE(g(Y_0,C)|A=1,Z,C)+(1-A)g(Y_0,C)\biggr\}\biggr]\\
%&=&E[\bigl(\omega(Z,C)-E(\omega(Z,C)|C)\bigr)\{AE(g(Y_0,C)|A=1,Z,C)+(1-A)g(Y,C)\}]\\
&=&E[\bigl\{\omega(Z,C)-E(\omega(Z,C)|C)\bigr\}\{\Pr(A=1|Z,C)E(g(Y_0,C)|A=1,Z,C)\\
&&\quad\quad+\Pr(A=0|Z,C)E(g(Y_0,C)|A=0,Z,C)\}]\\
&=&E[\bigl\{\omega(Z,C)-E(\omega(Z,C)|C)\bigr\}E(g(Y_0,C)|Z,C)]\\
&=&E[\bigl\{E(\omega(Z,C)|C)-E(\omega(Z,C)|C)\bigr\}E(g(Y_0,C)|C)]\\
&=&0.
\end{eqnarray*}
Also,
\begin{eqnarray*}
&&\hat{\psi}^{DR}\\
&\xrightarrow{p}&E\biggl[\frac{1-A}{\Pr(A=1)}\frac{\pi(Y_0,Z,C)}{1-\pi(Y_0,Z,C)}\left\{Y_0-\frac{E(Y_0\exp(\alpha(Y_0,Z,C))|A=0,Z,C)}{E(\exp(\alpha(Y_0,Z,C))|A=0,Z,C)}\right\}\\
&&+\frac{AE(Y_0|A=1,Z,C)}{\Pr(A=1)}\biggr]\\
&=&E\biggl[\frac{1-A}{\Pr(A=1)}\exp\{\alpha(Y_0,Z,C)+\beta^{\ast}(Z,C)\}\biggl\{Y_0-\frac{E(Y_0\exp(\alpha(Y_0,Z,C))|A=0,Z,C)}{E(\exp(\alpha(Y_0,Z,C))|A=0,Z,C)}\biggr\}\\
&&+\frac{E(Y_0A)}{\Pr(A=1)}\biggr]\\
&=&E(\frac{AY_0}{\Pr(A=1)})\\
&=&E(Y_0|A=1)=\psi.
\end{eqnarray*}

\noindent Thus, $\hat\eta^{DR}$ and $\hat\psi^{DR}$ are consistent if the parametric models lie in $\mathcal{M}_y$.
 Therefore, $\hat\eta^{DR}$ and $\hat\psi^{DR}$ are DR for $\eta$ and $\psi$ respectively.
%\end{proof}

%~\ref{sec: identification}

\section{\bf Proofs for examples in Section 3}\label{sec: appendix_identi}
\noindent\textbf{Proof of example \ref{eg: saturate_logit}}
%\begin{proof}

Let $\Pr(A=0|Y_0,Z;\theta_1,\theta_2,\eta_1,\eta_2)=\text{expit}(\theta_1+\theta_2Z+\eta_1Y_0+\eta_2 Y_0Z)$ and $\Pr(Y_0=1;\tau)=\exp(\tau )$. We show that for any $(\theta_1,\theta_2,\eta_1,\eta_2,\tau)$, there exists $(\tilde\theta_1,\tilde\theta_2,\tilde\eta_1,\tilde\eta_2,\tilde\tau)\neq(\theta_1,\theta_2,\eta_1,\eta_2,\tau)$ such that 
\begin{equation}\label{eq: eq_sat}
\Pr(A=0|Y_0,Z;\theta_1,\theta_2,\eta_1,\eta_2)\Pr(Y_0;\tau)=\Pr(A=0|Y_0,Z;\tilde\theta_1,\tilde\theta_2,\tilde\eta_1,\tilde\eta_2)\Pr(Y_0;\tilde\tau).
\end{equation} Suppose there exists $\rho_1\neq0$ such that $\Pr(Y_0=0;\tilde\tau)/\Pr(Y_0=0;\tau)=\exp(\rho_1)$, thus, \eqref{eq: eq_sat} is equivalent to

%\vspace{-2mm}
\begin{equation}\label{eq: proof_sat}
\frac{\Pr(A=0|Y_0,Z;\theta_1,\theta_2,\eta_1,\eta_2)}{\Pr(A=0|Y_0,Z;\tilde\theta_1,\tilde\theta_2,\tilde\eta_1,\tilde\eta_2)}=\frac{\Pr(Y_0;\tilde\tau)}{\Pr(Y_0;\tau)}=\exp(\rho_1+\rho_2Y_0),
\end{equation} 

\noindent where $\rho_2=\log[\exp(-\rho_1-\tau)+\{\exp(\tau)-1\}/\exp(\tau)]$. 

Note that two different sets of parameters would lead to the same observed data distribution by properly choosing $\rho_1$ and choosing $\tilde\theta_1=\theta_1-\rho_1-\log\varpi_1$, $\tilde\theta_2=\theta_2+\log\varpi_1-\log\varpi_2$, $\tilde\eta_1=\eta_1-\rho_2+\log\varpi_1-\log\varpi_3$, $\tilde\eta_2=\eta_2+\log\varpi_2+\log\varpi_3-\log\varpi_1-\log\varpi_4$ and $\tilde\tau=\tau+\rho_1+\rho_2$, where $\varpi_1=1+\exp(\theta_1)-\exp(\theta_1-\rho_1)$, $\varpi_2=1+\exp(\theta_1+\theta_2)-\exp(\theta_1+\theta_2-\rho_1)$, $\varpi_3=1+\exp(\theta_1+\eta_1)-\exp(\theta_1+\eta_1-\rho_1-\rho_2)$ and $\varpi_4=1+\exp(\theta_1+\theta_2+\eta_1+\eta_2)-\exp(\theta_1+\theta_2+\eta_1+\eta_2-\rho_1-\rho_2)$. For example, choose $\rho_1=0.3$, $\rho_2=-0.38$, $(\theta_1,\theta_2,\eta_1,\eta_2,\tau_1)=(0.3,0.6,0.1,0.7,-0.2)$ and $(\tilde\theta_1,\tilde\theta_2,\tilde\eta_1,\tilde\eta_2,\tilde\tau)=(-0.3,0.41,0.91,1.37,-0.28)$, they lead to the same observed distribution.
%\end{proof}

\noindent\textbf{Proof of example \ref{eg: sep_logit_bin}}

The separable treatment mechanism implies $\eta_2=\tilde\eta_2=0$, and thus $\varpi_2\varpi_3=\varpi_1\varpi_4$, i.e. $\{1+\exp(\theta_1+\theta_2)-\exp(\theta_1+\theta_2-\rho_1)\}
\{1+\exp(\theta_1+\eta_1)-\exp(\theta_1+\eta_1-\rho_1-\rho_2)\}=\{1+\exp(\theta_1)-\exp(\theta_1-\rho_1)\}\{1+\exp(\theta_1+\theta_2+\eta_1)-\exp(\theta_1+\theta_2+\eta_1-\rho_1-\rho_2)\}$
which indicates 
\begin{equation}\label{eq: sat_relation}
\exp(\rho_2)=\frac{\exp(\eta_1)}{1+\exp(\eta_1+\rho_1)-\exp(\rho_1)}.
\end{equation} Since in \eqref{eq: proof_sat}, $\exp(\rho_1+\rho_2Y_0)$ is the ratio of two densities for $Y_0$, we have $\rho_1$ and $\rho_1+\rho_2$ should be of the opposite sign. From equation \eqref{eq: sat_relation}, if $\rho_1>0$, then $\exp(\rho_1)>1$ and $\exp(\rho_1+\rho_2)>1$. Similarly, if $\rho_1<0$, then $\exp(\rho_1)<1$ and $\exp(\rho_1+\rho_2)<1$. Thus, we conclude that $\rho_1=\rho_2=0$, i.e. the separable treatment mechanism is identified for binary case.

\noindent\textbf{Proof of example \ref{eg: sep}}
%\begin{proof}

Suppose there exist two densities that make the ratios equal, 
\begin{eqnarray}\label{eq:logit0}
\frac{\text{expit}\{q_1(Z) + h_1(Y_0)\}}{\text{expit}\{q_2(Z) + h_2(Y_0)\}}=\frac{f_2(Y_0)}{f_1(Y_0)}.
\end{eqnarray}
We first take derivatives over $Z$ on both sides, and we have 
\[\frac{\partial \text{expit}\{q_1(Z)+h_1(Y_0)\}/\partial Z}{\text{expit}\{q_1(Z)+h_1(Y_0)\}}=\frac{\partial \text{expit}\{q_2(Z)+h_2(Y_0)\}/\partial Z}{\text{expit}\{q_2(Z)+h_2(Y_0)\}}.\]
Expand the $\text{expit}$ functions and simplify the equation yields
\begin{eqnarray}
&&\label{eq:logit}\\
&&\frac{\partial q_1(Z)/\partial Z}{\partial q_2(Z)/\partial Z}[1+\exp\{q_2(Z)+h_2(Y_0)\}]=1+\exp\{q_1(Z)+h_1(Y_0)\}.\nonumber
\end{eqnarray}
Next, we take derivatives over $Y_0$ on both sides of the above equation, and we have
\[\frac{\partial q_1(Z)/\partial Z}{\partial q_2(Z)/\partial Z}\frac{\partial h_2(Y_0)}{\partial Y_0}\exp\{q_2(Z)+h_2(Y_0)\}=\frac{\partial h_1(Y_0)}{\partial Y_0}\exp\{q_1(Z)+h_1(Y_0)\},\]
which is equivalent to,
\[\frac{\partial q_1(Z)/\partial Z}{\partial q_2(Z)/\partial Z}\exp\{q_2(Z)-q_1(Z)\}=\frac{\partial h_1(Y_0)/\partial Y_0}{\partial h_2(Y_0)/\partial Y_0}\exp\{h_1(Y_0)-h_2(Y_0)\}.\]
The left hand side of the above equation is a function of $Z$, but the right hand side is a function of $Y_0$. Thus, we must have 
\[\frac{\partial q_1(Z)/\partial Z}{\partial q_2(Z)/\partial Z}\exp\{q_2(Z)-q_1(Z)\}=c_1,\]
for some constant $c_1$. We multiply both sides of equation (\ref{eq:logit}) by $\exp\{-q_1(Z)\}$, and we have 
\[c_1[\exp\{-q_2(Z)\}+\exp\{h_2(Y_0)\}]=\exp\{-q_1(Z)\}+\exp\{h_1(Y_0)\},\]
and thus for some constant $c_2$,
\[c_1\exp\{-q_2(Z)\} +c_2=\exp\{-q_1(Z)\},\quad c_1\exp\{h_2(Y_0)\}-c_2=\exp\{h_1(Y_0)\}.\]
We substitute $q_2(Z)$ and $h_2(Y_0)$ in equation (\ref{eq:logit0}) with the expressions above to obtain
\[\exp\{h_1(Y_0)\}+c_2=\exp\{h_1(Y_0)\}\frac{f_1(Y_0)}{f_2(Y_0)},\]
and thus 
\[\frac{f_1(Y_0)}{f_2(Y_0)}=1+c_2\exp\{-h_1(Y_0)\}.\]
Note that $1+c_2\exp\{-h_1(Y_0)\}>1$ for $c_2>0$, and $1+c_2\exp\{-h_1(Y_0)\}<1$ for $c_2<0$. This cannot be true for the ratio of two densities. So we must have $c_2=0$, and thus $f_1(Y_0)/f_2(Y_0)=1$.
As a result, the joint distribution is identified.
%\end{proof}

The joint distribution is also identified in the separable treatment mechanisms for continuous outcome with binary instrument.
\begin{example}\label{eg: sep_binIV_contY}
Consider the case of continuous outcome with binary instrument.
Assume the Logistic separable treatment mechanism: $\mathcal{P}_{A|Y_0,Z}=\{\Pr(A=0|Y_0,Z):\logit \Pr(A=0|Y_0,Z) = \theta Z + h(Y_0)\}$, where $h$ is a known or unknown function. It can be shown that $\mathcal{P}_{A|Y_0,Z}$ satisfies the condition \ref{cond: not1} and thus the joint distribution is identified. 
\end{example}

\noindent\textbf{Proof of example \ref{eg: sep_binIV_contY}}
%\begin{proof}

Suppose there exist two sets of densities make the ratios equal, 
\begin{eqnarray}\label{eq:logit01}
\frac{\text{expit}\{\theta_1 Z + h_1(Y_0)\}}{\text{expit}\{\theta_2Z + h_2(Y_0)\}}=\frac{f_2(Y_0)}{f_1(Y_0)}.
\end{eqnarray}
The above equation holds for both $Z=0,1$, so we have 
\[\frac{ \text{expit}\{h_1(Y_0)\} }{\text{expit}\{h_2(Y_0)\}}=\frac{ \text{expit}\{\theta_1+h_1(Y_0)\}}{\text{expit}\{\theta_2+h_2(Y_0)\}}.\]
Simplifying the equation, we have 
\[\exp\{h_1(Y_0)\}=\frac{\exp(\theta_2)-\exp(\theta_1)+\{\exp(\theta_2)-\exp(\theta_1+\theta_2)\}\exp\{h_2(Y_0)\}}{\exp(\theta_1)-\exp(\theta_1+\theta_2)}.\]
Substituting $\exp\{h_1(Y_0)\}$ with the above expression in equation \eqref{eq:logit01}, we have 
\[\frac{f_2(Y_0)}{f_1(Y_0)}=1+\frac{\exp(\theta_2)-\exp(\theta_1)}{\exp(\theta_2)-\exp(\theta_1+\theta_2)}\exp\{-h_2(Y_0)\}.\]
If $\theta_1\neq \theta_2$, we must have $f_2(Y_0)/f_1(Y_0)<1$ for any $Y_0$, or $f_2(Y_0)/f_1(Y_0)>1$ for any $Y_0$. 
This cannot be true for the ratio of two densities. So we must have $\theta_1= \theta_2$, and thus $f_1(Y_0)/f_2(Y_0)=1$.
As a result, the joint distribution is identified.
%\end{proof}

\begin{example}\label{eg: psep}
Assume the Probit separable treatment mechanism: $\mathcal{P}_{A|Y_0,Z}\\=\{\Pr(A=0|Y_0,Z): \Pr(A=0|Y_0,Z) = \Phi\{q(Z) + h(Y_0)\}\}$, where $\Phi$ is the standard normal distribution function, $q$ and $h$ are known or unknown functions, and $q$ is differentiable. Then the joint distribution of $A,Y_0,Z$ is identified. 
\end{example}

\noindent\textbf{Proof of example \ref{eg: psep}}
%\begin{proof}

Suppose two sets of parameters make the ratio being a function of $Y_0$, i.e. for some function $s$,
\[\Phi\{q_1(Z) + h_1(Y_0)\}=\Phi\{q_2(Z) + h_2(Y_0)\}s(Y_0).\]
By taking derivatives over $Z$ on both sides, we have 
\[ \frac{\partial q_1(Z)}{\partial Z}\phi\{q_1(Z)+h_1(Y_0)\}= \frac{\partial q_2(Z)}{\partial Z}\phi\{q_2(Z) + h_2(Y_0)\}s(Y_0),\]
where $\phi$ is the standard normal density function. And equivalently
\[\log\frac{\phi\{q_1(Z)+h_1(Y_0)\}}{\phi\{q_2(Z)+h_2(Y_0)\}}= \log \frac{\partial q_2(Z)/\partial Z}{\partial q_1(Z)/\partial Z} + \log s(Y_0),\]
which implies that 
\begin{eqnarray}
&&\label{eq:probit}\\
\{q_2(Z)+h_2(Y_0)\}^2-\{q_1(Z)+h_1(Y_0)\}^2=2\left\{\log \frac{\partial q_2(Z)/\partial Z}{\partial q_1(Z)/\partial Z} + \log s(Y_0)\right\}.\nonumber
\end{eqnarray}
Note that the right hand side does not include an interaction term of $Z$ and $Y_0$, we have 
\[q_1(Z)h_1(Y_0)=q_2(Z)h_2(Y_0),\]
and thus
\[\frac{q_1(Z)}{q_2(Z)}=\frac{h_2(Y_0)}{h_1(Y_0)}.\]
Hence $q_1(Z)=cq_2(Z)$ and $h_2(Y_0)=ch_1(Y_0)$ for some positive constant $c$. Substituting $q_2$ and $h_2$ with $1/cq_1$ and $ch_1$ in equation (\ref{eq:probit}), we have 
\[\left(\frac{1}{c^2}-1\right)q_1^2(Z)+(c^2-1)h_1(Y_0)^2=2\{-\log c+\log s(Y_0)\}.\]
Since the right hand side does not vary with $Z$, we must have $c=1$, and thus $q_1(Z)=q_2(Z)$ and $h_2(Y_0)=h_1(Y_0)$. 
%\end{proof}

\section{\bf Additional results mentioned in this paper}\label{sec: appendix_more_deri}
\textbf{Regression estimator using any link function $\lambda$}

Let $\delta(Z,C)=\lambda\{ E(Y_0|A=0,Z,C)\}$ and $\tilde\alpha(A,Z,C)=\lambda\{ E(Y_0|A,Z,C)\}-\lambda\{ E(Y_0|A=0,Z,C)\}$, then $E(Y_0|A=1,Z,C)=\lambda^{-1} \{\tilde\alpha(1,Z,C)+\delta(Z,C)\}$. Let $\delta(Z,C;\xi)$ denote a parametric model for $\delta(Z,C)$ and let $\hat\xi$ denote the restricted MLE of $\xi$ using only data among the unexposed. Although in the main text $\eta$ is used to denote the parameter in $\alpha$, here we use it to denote the parameters in $\tilde\alpha$. We obtain an estimator for $\eta$ by solving:

\begin{equation}\label{eq: reg_alpha_anylink}
%\mathbb{P}_n\biggl[\bigl\{w(Z,C)- E(w(Z,C)|C;\hat\rho)\bigr\}\biggl\{A\frac{ E[\exp\{\hat\alpha(Y,Z,C;\eta)\}g(Y,C)|A=0,Z,C;\hat\xi]}{ E[\exp\{\hat\alpha(Y,Z,C;\eta)\}|A=0,Z,C;\hat\xi]}+(1-A)g(Y,C)\biggr\}\biggr]=0,
\mathbb{P}_n\biggl[\bigl\{w(Z,C)- E(w(Z,C)|C;\hat\rho)\bigr\}\biggl\{A\lambda^{-1} \{\tilde\alpha(1,Z,C;\eta)+\delta(Z,C;\hat\xi)\}+(1-A)Y\biggr\}\biggr]=0,
\end{equation}

We have the following proposition for the outcome regression estimator with any link function $\lambda$, the proof is similar to that of Proposition \ref{prop: reg_unbias}.

\begin{proposition}\label{prop: reg_unbias_anylink}
Under (IV.1)--(IV.2) and condition \ref{cond: not1}, suppose that $\tilde\alpha(A,Z,C;\eta)$, $f_{Z|C}(z|c;\rho)$ and $\delta(Z,C;\xi)$ are correctly specified, then the outcome regression estimator 
\begin{equation*}
\hat{\psi}^{reg}=\mathbb{P}_n\frac{A}{\hat\Pr(A=1)}\lambda^{-1} \{\tilde\alpha(1,Z,C;\hat\eta)+\delta(Z,C;\hat\xi)\},
\end{equation*}
is consistent for $\psi$.
\end{proposition}

\noindent\textbf{Equation \eqref{eq: Y0_correct} contains the correct model for $E(Y|A=0,Z,C)$}

%We verify that equation \eqref{eq: Y0_correct} for $E(Y|A=0,Z,C;\xi)$ contains the true data generating mechanism, thus it suffices to use \eqref{eq: Y0_correct} as a correctly specified outcome regression model in the simulations for binary outcome.

%The following result was used in the simulations for binary outcome. For binary $Y$, we derive the relationship between the regression model $\Pr(Y_0=1|A=1,Z,C)$ and data generating model $\Pr(Y_0=1|C)$ and this casts light on how to control the degree of misspecification of the regression model through the data generation model:

Since 

\vspace{-2mm}
\begin{eqnarray*}
&&\logit \Pr(Y_0=1|A,Z,C)\\
&=&\log \frac{\Pr(Y_0=1|A,Z,C)}{\Pr(Y_0=0|A,Z,C)}\\
&=&\log\Bigl\{ \frac{\Pr(Y_0=1|A,Z,C)}{\Pr(Y_0=0|A,Z,C)}/ \frac{\Pr(Y_0=1|A=0,Z,C)}{\Pr(Y_0=0|A=0,Z,C)}\Bigr\}\\
&&-\log\Bigl\{ \frac{\Pr(Y_0=1|Z,C)}{\Pr(Y_0=0|Z,C)}/ \frac{\Pr(Y_0=1|A=0,Z,C)}{\Pr(Y_0=0|A=0,Z,C)}\Bigr\}+\log\frac{\Pr(Y_0=1|Z,C)}{\Pr(Y_0=0|Z,C)},
\end{eqnarray*}

\vspace{-2mm}
\noindent and 
 
\vspace{-4mm}
\begin{eqnarray*}
&&\sum_a \frac{\Pr(Y_0=1|A=a,Z,C)}{\Pr(Y_0=0|A=a,Z,C)}\Pr(A=a|Y_0=0,Z,C)\\ 
&=&\sum_a \frac{\Pr(Y_0=1,A=a|Z,C)}{\Pr(Y_0=0,A=a|Z,C)}\Pr(A=a|Y_0=0,Z,C)\\ 
&=&\sum_a \frac{\Pr(Y_0=1,A=a|Z,C)}{\Pr(Y_0=0|Z,C)\Pr(A=a|Y_0=0,Z,C)}\Pr(A=a|Y_0=0,Z,C)\\ 
&=&\sum_a \frac{\Pr(Y_0=1,A=a|Z,C)}{\Pr(Y_0=0|Z,C)}= \frac{\Pr(Y_0=1|Z,C)}{\Pr(Y_0=0|Z,C)},
\end{eqnarray*}

%i.e. we can marginalize the ratio $\Pr(Y_0=1|A,Z,C)/\Pr(Y_0=0|A,Z,C)$ using the probability $\Pr(A|Y_0=0,Z,C)$ to get the marginalized ratio $\Pr(Y_0=1|Z,C)/\Pr(Y_0=0|Z,C)$. Thus,
\vspace{-2mm}
\noindent we have

\vspace{-2mm}
\begin{eqnarray*}
&&\logit \Pr(Y_0=1|A,Z,C)\\
&=&\log\Bigl\{ \frac{\Pr(Y_0=1|A,Z,C)}{\Pr(Y_0=0|A,Z,C)}/ \frac{\Pr(Y_0=1|A=0,Z,C)}{\Pr(Y_0=0|A=0,Z,C)}\Bigr\}\\
&&-\log\Bigl\{ \frac{\Pr(Y_0=1|Z,C)}{\Pr(Y_0=0|Z,C)}/ \frac{\Pr(Y_0=1|A=0,Z,C)}{\Pr(Y_0=0|A=0,Z,C)}\Bigr\}+\log\frac{\Pr(Y_0=1|Z,C)}{\Pr(Y_0=0|Z,C)}\\
&=&\log\Bigl\{ \frac{\Pr(Y_0=1|A,Z,C)}{\Pr(Y_0=0|A,Z,C)}/ \frac{\Pr(Y_0=1|A=0,Z,C)}{\Pr(Y_0=0|A=0,Z,C)}\Bigr\}\\
&&-\log\Bigl\{\sum_a \frac{\Pr(Y_0=1|A=a,Z,C)}{\Pr(Y_0=0|A=a,Z,C)}/ \frac{\Pr(Y_0=1|A=0,Z,C)}{\Pr(Y_0=0|A=0,Z,C)}\Pr(A=a|Y_0=0,Z,C)\Bigr\}\\
&&+\log\frac{\Pr(Y_0=1|Z,C)}{\Pr(Y_0=0|Z,C)}\\
&=&\alpha(1,Z,C)A-\log\biggl[\exp\{\alpha(1,Z,C)\}\Pr(A=1|Y_0=0,Z,C)\\
&&+\Pr(A=0|Y_0=0,Z,C)\biggr]+\logit \Pr(Y_0=1|C).
\end{eqnarray*}
In our simulation $\alpha(1,Z,C)=\eta$ and $\Pr(A|Y_0,Z,C)=\Pr(A|Y_0,Z,C_1)$, thus
\begin{eqnarray*}
&&\logit \Pr(Y_0=1|A,Z,C)\\
&=&\alpha(1,Z,C)A-g(Z,C_1)+\logit \Pr(Y_0=1|C_1,C_2),
\end{eqnarray*}
where $g(Z,C_1)=\log\{\exp\{\alpha(1,Z,C)\}\Pr(A=1|Y_0=0,Z,C)+\Pr(A=0|Y_0=0,Z,C)\}$. Since $\logit\Pr(Y_0=1|C)$ is linear in $C_2$, so does $\logit \Pr(Y_0=1|A,Z,C)$.%Thus we can control the effect of $C_2$ in the model $\Pr(Y_0=1|A,Z,C)$ through $\Pr(Y_0=1|C)$.

%\noindent\textbf{Characterization of orthogonal tangent space in (ii)}

%The notations in the following derivation are inconsistent with other notations in the rest of the paper.
%{\color{blue}

\section{\bf Local efficiency}\label{sec: appendix_efficiency}

Let $\left( A,L\right) =(A,Y_0,Z,C)$ and $O=(A,Y,Z,C)$ denote the full data and observed data respectively. Assume $\logit \pi(Y_0,Z,C)=\alpha(Y_0,Z,C)+\beta(Z,C)$, where $\beta(Z,C)$ is unrestricted, $\alpha(Y_0,Z,C)$ is known and assume $Y_0\indep Z|C$. First, we derive the observed data orthogonal tangent space. All the scores of $f(Y_0,Z,C)$ can be written as 

\begin{equation*}
\mathcal{N}_1=\{S(Y_0,Z,C):S(Y_0,Z,C)=S(Y_0|C)+S(Z|C)+S(C)\}. 
\end{equation*} 

\noindent where $E\{S(Y_0|C)|C\}=E\{S(Z|C)|C\}=E\{S(C)\}=0$. Therefore, by \citet{bickel1998efficient} and \citet{tchetgen2010doubly}, we can show that 
\[\mathcal{N}_1^{\perp}=\{v(Y_0,Z,C)-v^{\dagger}(Y_0,Z,C): \text{ for any }v(Y_0,Z,C)\}.\] 

\noindent where $v^{\dagger}(Y_0,Z,C;\phi)=E(v|Z,C;\phi)+E(v|Y_0,C;\phi)-E(v|C;\phi).$ Let $\mathcal{N}_2$ denote the tangent space of all the treatment propensity score. Thus, we have \[\mathcal{N}_{2}=\biggl\{ \left\{ A-\pi \left( L\right)
\right\} u(Z,C)\text{ for all }u\biggr\}.\] Therefore, the tangent space for the full data $(A,L)$ is $\mathcal {N}=\mathcal{N}_1\bigoplus\mathcal{N}_2$, where $\bigoplus$ denotes direct summation of spaces. \citet{rotnitzky1997analysis} showed that the observed data tangent space is given by $%A=1
\mathcal{N}^{O}=\overline{\mathcal{N}_{1}^{O}+\mathcal{N}_{2}^{O}},$ where $%
\mathcal{N}_{j}^{O}=\overline{R\left( g\circ \Pi _{j}\right) }$, $%
R\left( \cdot \right) $ is the range of the operator $g:\Omega ^{\left(
A,L\right) }\rightarrow \Omega ^{\left( O\right) }$ and $g$ is the conditional
expectation operator $g\left( \cdot \right) =E\left[ \cdot |O\right] ,$ $%
\Omega ^{\left( A,L\right) }$ and $\Omega ^{\left( O\right) }$ are the
spaces of all random functions of $(A,L)$ and $O$ respectively. $\Pi _{j}$
is the Hilbert space projection operator from $\Omega ^{\left( A,L\right) }$
onto $\mathcal{N}_{j}$ and $\overline{\mathcal{S}}$ is the close linear span
of the set $\mathcal{S}$.

As shown in \citet{bickel1998efficient}, the orthocomplement to the
tangent space in the observed data model $\mathcal{N}^{O,\bot }=\mathcal{N}%
_{1}^{O,\bot }\cap \mathcal{N}_{2}^{O,\bot }$. \citet{rotnitzky1997analysis} established that 

\begin{eqnarray*}
\begin{split}
\mathcal{N}_{1}^{O,\bot } &=&
\left\{ \frac{1-A}{1-\pi \left(
L\right)}m(L)+N_{car}:m\left( L\right) \in \mathcal{N}_{1}^{\bot }\text{ and }%
N_{car}\in \mathcal{N}_{car}\text{ }\right\},\\
 \text{where }&& \\
\mathcal{N}_{car} &=&\left\{ Ak\left( O\right) -\frac{1-A}{1-\pi \left( L\right)}E\left[
Ak\left( O\right) |L\right] :\text{ for any }%
k\left( O\right) \text{ }\in \Omega ^{\left( O\right) }\right\} .
\end{split}\end{eqnarray*}%

\noindent Therefore, by the formula of $\mathcal{N}_{1}$, we have $\mathcal{N}_{1}^{O,\bot }$ consists of functions 
\[
\frac{1-A}{1-\pi\left( L\right)}\left\{ v-v^{\dag }\right\}   +Ak\left( O\right) -\frac{1-A}{1-\pi \left( L\right)}E
\left[ Ak\left( O\right) |L\right]. 
\]%

Also, Rotnitzky and Robins
(1997) establish that $\mathcal{N}_{2}^{O,\bot }=\left\{ b\left( O\right)
:b\left( O\right) \in \mathcal{N}_{2}^{\bot }\right\} $. Therefore, $\mathcal{N}^{O,\bot }=\left\{ N_{1}^{O,\bot }\in \mathcal{N}_{1}^{O,\bot }:E%
\left[ N_{2}N_{1}^{O,\bot }\right] =0,N_{2}\in \mathcal{N}_{2}\right\}$. Thus, we have the following
result. 

\begin{lemma}\label{lemma: tangent_space_for_(ii)} 

We have
\[
\mathcal{N}^{O,\bot }=\left\{ 
\begin{array}{c}
\begin{split}
\frac{1-A}{1-\pi \left(
L\right)}\left\{ v-v^{\dag }\right\}  +Ak_{v}\left( O\right) -\frac{1-A}{1-\pi \left(
L\right)}E\left[ Ak_{v}\left( O\right) |L
\right]   : 
\end{split} \\
k_{v}=E\left[ v-v^{\dag }|A=1,Z,C\right] 
\end{array}%
\right\}. 
\]
\end{lemma}

\begin{proof}
$N_{1}^{O,\bot }\left( k_{v}\right) $ is clearly in $\mathcal{N}_{1}^{O,\bot
},$ it suffices to show that $N_{1}^{O,\bot ^{\ast }}=N_{1}^{O,\bot
}\left( k_{v}\right)$ is the unique solution to the equation $E\left[
N_{1}^{O,\bot ^{\ast }}N_{2}\right] =0,$ for all $v$ and for all $%
N_{2}\in \mathcal{N}_{2}$. In this vein,  
\begin{eqnarray*}
0 &=&E\left[ N_{1}^{O,\bot ^{\ast }}N_{2}\right]  \\
&=&E\left[ \left\{ 
\begin{array}{c}
(1-A)\left\{ v-v^{\dag }\right\} /\{1-\pi \left(
L\right)\}  +Ak^{\ast }\left(
O\right)  \\ 
-(1-A)E\left[ Ak^{\ast }\left( O\right) |L\right] /\{1-\pi \left(
L\right)\} 
\end{array}%
\right\} \left( A-\pi \left( L\right) \right) u(Z,C)\right]\text{ for all 
}u. \\
\end{eqnarray*}

This is equivalent to 
\begin{eqnarray*}
&&E\left[ \left\{ 
\begin{array}{c}
(1-A)\left\{ v-v^{\dag }\right\} /\{1-\pi \left(
L\right)\}  +Ak^{\ast }\left(
O\right)  \\ 
-(1-A)E\left[ Ak^{\ast }\left( O\right) |L\right] /\{1-\pi \left(
L\right)\}  
\end{array}%
\right\} \left( A-\pi \left( L\right) \right) |Z,C)\right]  =0 \\
&\Leftrightarrow & E\left[ \left( 1-\pi \left( L\right) \right) \left\{
v-v^{\dag }\right\} |Z,C\right] -E\left[ (1-\pi \left( L\right) )\pi \left(
L\right) k^{\ast }\left( O\right) |Z,C\right]  \\
&&-E\left[ \left( 1-\pi \left( L\right) \right) E\left[ Ak^{\ast }\left(
O\right) |L\right] |Z,C\right] =0\\
&\Leftrightarrow &E\left[ \left( 1-\pi \left( L\right) \right) \left\{
v-v^{\dag }\right\} |Z,C\right] -E\left[ \left( 1-\pi \left( L\right)
\right) k^{\ast }\left( O\right) |Z,C\right]  =0\\
&\Leftrightarrow &E\left[ \left[ E\left[ \left\{ v-v^{\dag }\right\}
|C,A=1,Z\right] -k^{\ast }\left( O\right) \right] A|Z,C\right] =0.
\end{eqnarray*}%
Upon writing $k^{\ast }\left( O\right) =k_{1}^{\ast }\left( L\right)
(1-A)+k_{2}^{\ast }\left( Z,C\right) A,$ we have that $k_{2}^{\ast }\left(
Z,C\right) =E\left[ \left\{ v-v^{\dag }\right\} |C,A=1,Z\right] =k_{v},$
proving the result of Lemma \ref{lemma: tangent_space_for_(ii)}. 
\end{proof}

Note that simple algebra yeilds that $\mathcal{N}^{O,\perp}=\mathcal{J}^{\perp}$, where 

\begin{eqnarray}\label{eq: obs_tangent_space_for_ii}
&&\\
\mathcal{J}^{\perp}&=&\biggl\{\frac{1-A}{1-\pi}(v-v^{\dagger})+\frac{A-\pi}{1-\pi}E(v-v^{\dagger}|A=1,Z,C;\gamma,\xi): \text{ any function }v=v(Y_0,Z,C)\biggr\}\nonumber\\
&=&\biggl\{
\begin{array}{c}
\tilde Q_{v-v^{\dagger}}(Y,A,Z,C;\gamma,\xi): \text{ any function }v=v(Y,Z,C)
\end{array}\biggr\},\nonumber
\end{eqnarray}

\noindent where $\pi=\pi(Y,Z,C)$. 

Heretofore, we have derived the orthogonal tangent space assuming the selection bias $\alpha(Y_0,Z,C)$ is known and $Y_0\indep Z|C$. We next show that assuming the selection bias function $\alpha(Y_0,Z,C)$ is correctly specified and $Y_0\indep Z|C$, the space of influence functions for all RAL estimators for the parameter of interest $\psi$ is as follows:

\begin{eqnarray}\label{eq: all_IF_psi}
&&\\
\mathcal{J_{\psi}}^{\perp}%&=&\biggl\{\frac{1-A}{1-\pi}(d+v-v^{\dagger})-\frac{A-\pi}{1-\pi}E(d+v-v^{\dagger}|A=1,Z,C): \text{ any function }v=v(Y_0,Z,C)\biggr\}\nonumber\\
&=&\biggl\{
\begin{array}{c}
\tilde Q_{d+v-v^{\dagger}}(Y,A,Z,C;\gamma,\xi)
-E[\bigtriangledown_{\eta} \tilde Q_{d+v-v^{\dagger}}(Y,A,Z,C;\gamma,\xi)|\gamma,\xi]IF_{\hat\eta}(\gamma,\xi,\phi):\\
 \text{ any function }v=v(Y,Z,C)
\end{array}\biggr\},\nonumber
\end{eqnarray}

\noindent where $d=d(A,Y_0;\psi)$ is any function proportional to $AY_0/\Pr(A=1)-\psi$ and $IF_{\hat\eta}$ is the influence function of any RAL estimator $\hat\eta$ for $\eta$ and $IF_{\hat\eta}$ belongs to $\mathcal{J}^{\perp}$. Note that the estimators given in Proposition \ref{prop: DR_unknown_alpha0} is a subset of all RAL estimators for $\psi$, their influence function also belongs to \eqref{eq: all_IF_psi}. More specifically, the influence functions of all estimators for $\psi$ given in Proposition \ref{prop: DR_unknown_alpha0} are:

\[\tilde Q_{d}(Y,A,Z,C;\gamma,\xi)
-E[\bigtriangledown_{\eta} \tilde Q_{d}(Y,A,Z,C;\gamma,\xi)|\gamma,\xi]IF_{\tilde \eta},\]

\noindent where $\tilde \eta$ is the DR estimator of $\eta$.

To derived the efficiency bound for all regular and asymptotically linear estimators for $\psi$, let $\Pi(\cdot|\mathcal{J}^{\perp})$ denote the projection onto the Hilbert space $\mathcal{J}^{\perp}$. We have the following result for the efficiency.

\begin{proposition}\label{prop: eff_IF}
Under (IV.1)-(IV.2) and condition \ref{cond: not1}, we have $\hat\psi^{eff}=Q_{\tilde g}(Y,A,Z,C;\gamma^{\dagger},\hat\xi)/\hat\Pr(A=1)-M$ is the efficient RAL estimator for $\psi$ with influence function

\[\frac{Q_{\tilde g}(Y,A,Z,C;\gamma,\xi)}{\Pr(A=1)}-M-\psi+E[\bigtriangledown_{\eta}(\frac{Q_{\tilde g}(Y,A,Z,C;\gamma,\xi)}{\Pr(A=1)}-M)]IF_{\eta^{\dagger}},\]

\noindent  where $\gamma^{\dagger}=(\eta^{\dagger},\tilde \theta)$, $\eta^{\dagger}$ is the most efficient estimator for $\eta$ and $M=\Pi\{Q_{\tilde g}(Y,A,Z,C;\gamma^{\dagger},\hat\xi)/\hat\Pr(A=1)|\mathcal{J}^{\perp}\}$. 
\end{proposition}

\begin{proof}
Recall $\logit{\pi}(Y_0,Z,C)=\alpha(Y_0,Z,C)+\beta(Z,C)$, where $\beta(Z,C)$ is unrestricted and $\alpha(0,Z,C)=0$. To derive the efficient influence function for $\psi$, we consider the following three model spaces:

(i) The selection bias $\alpha(Y_0,Z,C)$ is known.

(ii) The selection bias $\alpha(Y_0,Z,C)$ is known and $Y_0$ is independent with $Z$ conditional on $C$, i.e., $Y_0\indep Z|C$ holds.

(iii) The selection bias $\alpha(Y_0,Z,C)$ is parametrically specified as $\alpha(Y_0,Z,C;\gamma)$ and $Y_0$ is independent with $Z$ conditional on $C$, i.e., $Y_0\indep Z|C$, where $\gamma$ is unknown $p$-dimensional parameter.

Note that the tangent space of (i) is the entire Hilbert space $\mathcal{H}$ since there is no additional restriction on the joint data likelihood. \cite{robins2000sensitivity} showed that the efficient influence function for $\psi$ is $IF_{\psi,1}^{eff}=Q_{\tilde g}(Y,A,Z,C;\tilde\gamma,\hat\xi)/\hat\Pr(A=1)-\psi$ where $\tilde g(Y,C)=Y$. 

For (ii), we have shown that the observed data orthogonal tangent space is $\mathcal{J}^{\perp}$ as given in \eqref{eq: obs_tangent_space_for_ii}. Hence, the influence function for $\psi$ in (ii) is $IF_{\psi,2}=IF_{\psi,1}^{eff}+U(t)$ where $U(t)\in \mathcal{J}^{\perp}$ and $t$ is the parameters in a parametric submodel. 

For (iii), we first characterize the space of all influence functions for estimators for $\eta$, denoted as $IF_{\eta}$. Note that $\eta$ is variational independent of all the other nuisance parameters in (iii), i.e., $\eta$ is variational independent of all the parameters in (ii). Thus, the space for the influence functions of all RAL estimators for $\eta$ in (iii) is also ${\mathcal{J}^{\perp}}$. To derive the influence function for $\psi$ in (iii), note that $E_t\{IF_{\psi,2}(\psi(t),\eta(t),t)\}=0$, thus $\bigtriangledown_t E_t\{IF_{\psi,2}(\psi(t),\eta(t),t)\}=0$, where $\bigtriangledown$ is the Laplace operator. Also,

%\vspace{-8mm}
\begin{eqnarray*}
&&\bigtriangledown_t E_t\{IF_{\psi,2}(\psi(t),\eta(t),t)\}\\
&=&E_t\{IF_{\psi,2}(\psi(t),\eta(t),t)S_t(A,Y,Z,C)\}+E_t\{\bigtriangledown_{\psi}IF_{\psi,2}(\psi(t),\eta(t),t)\}\bigtriangledown_t \psi(t)\\
&&+E_t\{\bigtriangledown_{\eta}IF_{\psi,2}(\psi(t),\eta(t),t)\}\bigtriangledown_t \eta(t)+E_t\{\bigtriangledown_t  IF_{\psi,2}(\psi(t),\eta(t),t)\}.
\end{eqnarray*}

Note that 

\begin{eqnarray*}
IF_{\psi,2}(\psi(t),\eta(t),t)&=&IF_{\psi,1}^{eff}(\psi(t),\eta(t),t)+U(t)\\
&=&Q_{\tilde g}(Y,A,Z,C;t)/\Pr(A=1;t)-\psi+U(t),
\end{eqnarray*} 

\noindent thus $\bigtriangledown_{\psi}  IF_{\psi,2}(\psi(t),\eta(t),t)=-1$. Due to the robustness of $IF_{\psi,1}^{eff}(\psi(t),\eta(t),t)+\psi$ for $E(Y_0|A=1,Z,C)$, we have $E\{\bigtriangledown_t IF_{\psi,1}^{eff}(\psi(t),\eta(t),t)\}=0$. Similarly, the double robustness of $U(t)$ indicates that $E\{\bigtriangledown_t U(t)\}=0$. Thus, $E\{\bigtriangledown_t IF_{\psi,2}(\psi(t),\eta(t),t)\}=0$. 

Since,

\vspace{-3mm}
\begin{eqnarray*}
\frac{\partial \psi}{\partial t} &=&E_t\{IF_{\psi,2}(\psi(t),\eta(t),t)S_t(A,Y,Z,C)\}+E_t\{\bigtriangledown_{\eta}IF_{\psi,2}(\psi(t),\eta(t),t)\}\bigtriangledown_t \eta(t)\\
&=&E_t\biggl[\biggl\{IF_{\psi,2}(\psi(t),\eta(t),t)+E_t\{\bigtriangledown_{\eta}IF_{\psi,2}(\psi(t),\eta(t),t)\}IF_{\eta}\biggr\}S_t(A,Y,Z,C)\biggr],\\
\end{eqnarray*}

%\vspace{-8mm}
\noindent Thus,

 \[\biggl\{IF_{\psi,2}(\psi(t),\eta(t),t)+E_t\{\bigtriangledown_{\eta}IF_{\psi,2}(\psi(t),\eta(t),t)\}IF_{\eta}\biggr\}\] 
 
 \noindent is the space of influence functions for all RAL estimators for $\psi$ which is also the observed data orthocomplement to the nuisance tangent space for model (iii). Note that $IF_{\psi,2}(\psi(t),\eta(t),t)=IF_{\psi,1}^{eff}(\psi(t),\eta(t),t)+U(t)$, thus by choosing $U(t)=-\Pi\{IF_{\psi,1}^{eff}(\psi(t),\eta(t),t)|\mathcal{J}^{\perp}\}\in\mathcal{J}^{\perp}$ and $IF_{\eta}=IF_{\eta}^{eff}$, the influence function for an RAL estimator for $\psi$ is $\Pi(IF_{\psi,1}^{eff}(\psi(t),\eta(t),t)|\mathcal{J})+E[\bigtriangledown_{\eta}\{ \Pi(IF_{\psi,1}^{eff}(\psi(t),\eta(t),t)|\mathcal{J})\}]IF_{\eta}^{eff}$. Note that this influence function is in the tangent space of the model, thus it is the efficient influence function for $\psi$ in model (iii). That is,

%\vspace{-7mm}
\[IF_{\psi,3}^{eff}=\Pi(IF_{\psi,1}^{eff}(\psi(t),\eta(t),t)|\mathcal{J})+E[\bigtriangledown_{\eta}\{ \Pi(IF_{\psi,1}^{eff}(\psi(t),\eta(t),t)|\mathcal{J})\}]IF_{\eta}^{eff}.\]
\end{proof}

Proposition \ref{prop: eff_IF} provides a theorical efficiency bound for all regular estimators of ETT. Finding the optimal function $v$ such that $\eta^{\dagger}$ is the most efficient estimator might be challenging in practice. For illustration, we derive the efficient influence function for $\eta$ and $\psi$ where $Y$ and $Z$ are both binary. Thus, $v(Y_0,Z,C)$ can be written as  $v(Y_0,Z,C)=h_0(C)+h_1(C)Z+h_2(C)Y_0+h(C)Y_0Z$ and thus $v-v^{\dagger}=h(C)\bar v(Y_0,Z,C)$ where $ \bar v(Y_0,Z,C)=\{Y_0-E(Y_0|C)\}\{Z-E(Z|C)\}$. Let $U_h(t)=\tilde Q_{v-v^{\dagger}}(Y,A,Z,C;\gamma,\xi)$, thus, $U_h(t)=h(C)\Delta(t)$, where \[\Delta(t)=\tilde Q_{\bar v}(Y,A,Z,C;\gamma,\xi).\]

\noindent To find the efficient influence function of $\hat\psi^{DR}$, we first find the efficient influence function of $\eta$. Let ${h}^{opt}$ denote the choice of $h$ such that $\hat\eta$ is the most efficient. We have $U_{h}^{opt}$ satisfies $E[\partial U_{h}/\partial \eta]=E[U_{h}U_{h}^{optT}]$ for any $h$ (Newey and McFadden 1994, Chap 36\nocite{Newey1994large}). Thus, the efficient estimator for $\eta$ satisfies $\mathbb{P}_n[U_{h}^{opt}(\hat\eta^{eff})]=0$. That is \[E[h(C)\{\frac{\partial \Delta(t)}{\partial\eta^T}+\Delta(t)\Delta(t)^Th^{optT}(C)\}]=0.\] Select $h(C)=E[\partial \Delta(t)/\partial\eta^T+\Delta(t)\Delta(t)^Th^{optT}(C)|C]$, thus \[h^{opt}(C)=-E[\Delta(t)\Delta(t)^T|C]^{-1}E[\partial \Delta(t)/\partial\eta^T|C].\] Thus, the efficient influence function for $\eta$ is $IF_{\eta}^{eff}=h^{opt}(C)\Delta(t)$. Let $H=Q_{\tilde g}(Y,A,Z,C;\gamma^{\dagger},\hat\xi)/\hat\Pr(A=1)$ thus $M=\Pi\{H|\mathcal{J}^{\perp}\}$. Also note that $M\in\mathcal{J}^{\perp}$, thus $M$ could be written as $M=\tilde h(C)\Delta(t)$. Hence, we have $E[\{H-\tilde h(C)\Delta(t)\}h(C)\Delta(t)]=0$ for any $h(C)$. This is equivalent to $E[\{H-\tilde h(C)\Delta(t)\}\Delta(t)|C]=0$. Thus, we obtain $\tilde h(C)=E\{\Delta^2(t)|C\}^{-1}E\{H\Delta(t)|C\}$ which yields $M=E\{\Delta^2(t)|C\}^{-1}E\{H\Delta(t)|C\}\Delta(t)$ and by proposition \ref{prop: eff_IF}, we obtain the efficient influence function for $\psi$.

\clearpage
\newpage

%\section{\bf Validity of the identification assumptions}

\end{document}